\documentclass[aps,prb,groupedaddress,twocolumn,showpacs]{revtex4-1}
\usepackage{graphicx}
\usepackage{dcolumn}
\usepackage{bm}
\usepackage{amssymb}
\usepackage{epstopdf}
\usepackage{amsmath}
\usepackage{color}
\usepackage[dvipdfmx]{hyperref}
\hypersetup{
 setpagesize = false,
 colorlinks = true,
 citecolor = blue,
 linkcolor = blue,
 urlcolor = blue,
 }
\begin{document}
\title{Two-dimensional binary transition metal nitride $M$N$_4$ ($M$ = V, Cr, Mn, Fe, Co)
 with a graphene-like structure and strong magnetic properties}
\author{Shuo Zhang$^{1}$}\email{These authors contributed equally to this work.}
\author{Panjun Feng$^{1}$}\email{These authors contributed equally to this work.}
\author{Dapeng Liu$^{1}$}\email{These authors contributed equally to this work.}
\author{Hongfen Wu$^{1}$}
\author{Miao Gao$^{2}$}
\author{Tongshuai Xu$^{3}$}
\author{Z. Y. Xie$^{4}$}\email{Corresponding author: qingtaoxie@ruc.edu.cn}
\author{Xun-Wang Yan$^{1}$}\email{Corresponding author: yanxunwang@163.com}
\date{\today}
\affiliation{$^{1}$College of Physics and Engineering, Qufu Normal University, Qufu, Shandong 273165, China}
\affiliation{$^{2}$Department of Physics, School of Physical Science and Technology, Ningbo University, Zhejiang 315211, China}
\affiliation{$^{3}$ School of Physics and Electric Engineering, Anyang Normal University, Anyang 455002, China.}
\affiliation{$^{4}$Department of Physics, Renmin University of China, Beijing 100872, China.}

\begin{abstract}
  Binary transition metal nitride with a graphene-like structure and strong magnetic properties is rare. Based on the first-principles calculations, we design two kinds of $M$N$_4$ ($M$ =transition metal) monolayers, which are transition metal nitrides with a planar structure, made up of $M$N$_4$ units aligned in the rhombic and square patterns. The two structural lattices have robust stability and good compatibility with different metal atoms, and the underlying mechanism is the combination of $sp^2$ hybridization, coordinate bond, and $\pi$ conjugation. With the metal atom changing from V, Cr, Mn, Fe to Co, the total charge of $M$N$_4$ system increases by one electron in turn, which results in continuous adjustability of the electronic and magnetic properties. The planar ligand field is another feature of the two $M$N$_4$ lattices, which brings about the special splitting of five suborbitals of 3$d$ metal atom and gives rise to strong magnetism. Moreover, room-temperature ferromagnetism in square-CoN$_4$ monolayer with the Curie temperatures of 321 K is determined by solving the Heisenberg model combined with Monte Carlo method.
\end{abstract}

\maketitle

\section{Introduction}
Compared with metal oxides and chalcogenides, transition metal nitrides have distinctive electronic structure, resulting in different physical and chemical properties and various applications.
Bulk transition metal nitrides usually have the high hardness and high melting point, which are related to some applications including refractory material, cutting material, and hard coatings \cite{Salamat2013}.
Two-dimensional transition metal nitrides include MXenes Ti$_4$N$_3$, Mo$_2$N, and Ti$_2$N \cite{Zheng2020}, which can serve as energy storage and electrode materials due to superior conductivity and chemical stability.
Although these nitride MXenes are called two-dimensional materials, transition metal atoms are actually in the three-dimensional crystal field because nitride MXenes are composed of three or more layers of atoms.
For the single-atom-thick transition metal nitride, the unique electronic and magnetic properties can be expected because of its graphene-like structure and real two-dimensional crystal field.
However, this kind of material is very scarce in the family of transition metal nitrides.
From another point of view, the realization of intrinsic magnetism in graphene and graphene-like two-dimensional materials is a long-standing challenge in condensed matter physics and material fields \cite{Shabbir2018,Sethulakshmi2019}, which makes the exploration of single-atom-thick transition metal nitrides more significant.

In experiments, the planar $M$N$_4$ ($M$ = metal) moiety embedded in the graphene sheet acting as single atom catalyst has been identified \cite{Fei2018,He2019,Zhao2019}, and more than twenty different metals, including noble metals and non-noble metals, are employed to synthesize the $M$N$_4$-graphene structure. Recently, the room-temperature ferromagnetism in CoN$_4$-graphene systems has been observed \cite{Hu2021}.
More importantly, triclinic beryllium tetranitride BeN$_4$ was synthesized under the pressure of 85 GPa \cite{Bykov2021}, and it transforms to layered
van der Waals bonded BeN$_4$ with small exfoliation energy. The planar BeN$_4$ layer with the single-atomic thickness is a new class of 2D materials.
In theoretical studies, the planar cobalt carbonitrides CoN$_4$C$_{10}$, Co$_2$N$_8$C$_6$, Mn$_2$N$_6$C$_6$, CoN$_4$C$_2$, and CrN$_4$C$_2$ are predicted by the first-principles calculations \cite{Liu2021,Feng2022,Liu2021a,Liu2021b} and the topological electronic properties, such as topological quantum spin and valley Hall insulators, are investigated \cite{Wang2021,Dong2022}. In addition, single-atom-thick nitrides CrN, MnN$_2$, and MgN$_4$ are also reported \cite{Zhang2015,Zhao2020,Mortazavi2021}, which structural unit can be regarded as the planar $M$N$_4$ moiety with N atom being or not being shared by the neighboring units.
After the careful analysis of the above structures, we are surprised to find that $M$N$_4$ unit plays a vital role in keeping the planar geometry.
The finding gives us an inspiration that the $M$N$_4$ moiety may be used to build some free-standing monolayers of transition metal nitrides.

In this paper, we design two kinds of lattices for single-atom-thick transition metal nitrides, named as rhombus $M$N$_4$ ($r$-$M$N$_4$)and square $M$N$_4$ ($s$-$M$N$_4$).  The stability mechanism of the planar pattern is analyzed by the cooperation of $\pi$-d conjugation between metal and N atoms and $\pi$-$\pi$ conjugation in the polymeric nitrogen chain or square nitrogen ring. The amazing feature is that the two lattices are compatible with various metal elements from V, Cr, Mn, Fe, to Co, which results in the fact that the electron number of these single-atom-thick systems can be adjusted continually.
With the metal atoms varying, the planar metal nitrides display rich magnetic properties including the transition from antiferromagnetism to ferromagnetism and the evolution of magnetic moment.
The Curie temperature of $s$-CoN$_4$ ferromagnet is 321 K, which is an example to indicate that these planar transition metal nitrides have abundant magnetic properties.

\section{Method}
The plane wave pseudopotential method within VASP code and the projector augmented-wave (PAW) pseudopotential with Perdew-Burke-Ernzerhof (PBE) exchange-correlation functional \cite{PhysRevB.47.558, PhysRevB.54.11169, PhysRevLett.77.3865, PhysRevB.50.17953} were adopted in our calculations.
The plane wave basis cutoff is 600 eV and the thresholds are 10$^{-5}$ eV and 0.001 eV/\AA ~ for total energy and force convergence.
The interlayer distance was set to 20 \AA~ and a mesh of $24\times 24\times 1$ k-points was used for the Brillouin zone integration.
The phonon calculations are carried out with the supercell method in the PHONOPY program,
and the real-space force constants of supercells were calculated using density-functional perturbation theory (DFPT) as implemented in VASP  \cite{Togo2015}.
The force convergence criterion 10$^{-5}$ eV/\AA was used in structural optimization of the primitive cell before building the supercell.
In the ab initio molecular dynamics simulations,
the 3 $\times$ 3 $\times$ 1 supercells were employed and the temperature was kept at 1000 K for 5 ps with a time step of 1 fs in the canonical ensemble (NVT) \cite{Martyna1992}.
 The phase transition temperature is evaluated by the Monte Carlo method enclosed in the software package developed by Yehui Zhang et al.\cite{Zhang2021}, and the 100 $\times$ 100 $\times$ 1 lattice is used in the Monte Carlo simulation.
To get the more accurate the energies of $s$-CoN$_4$ monolayer in several magnetic orders, the nonempirical strongly constrained and appropriately normed (SCAN) meta-GGA method is employed \cite{Sun2015}.

\section{Results and discussion}

\subsection{Atomic structure}

\begin{figure}
\begin{center}
\includegraphics[width=7.0cm]{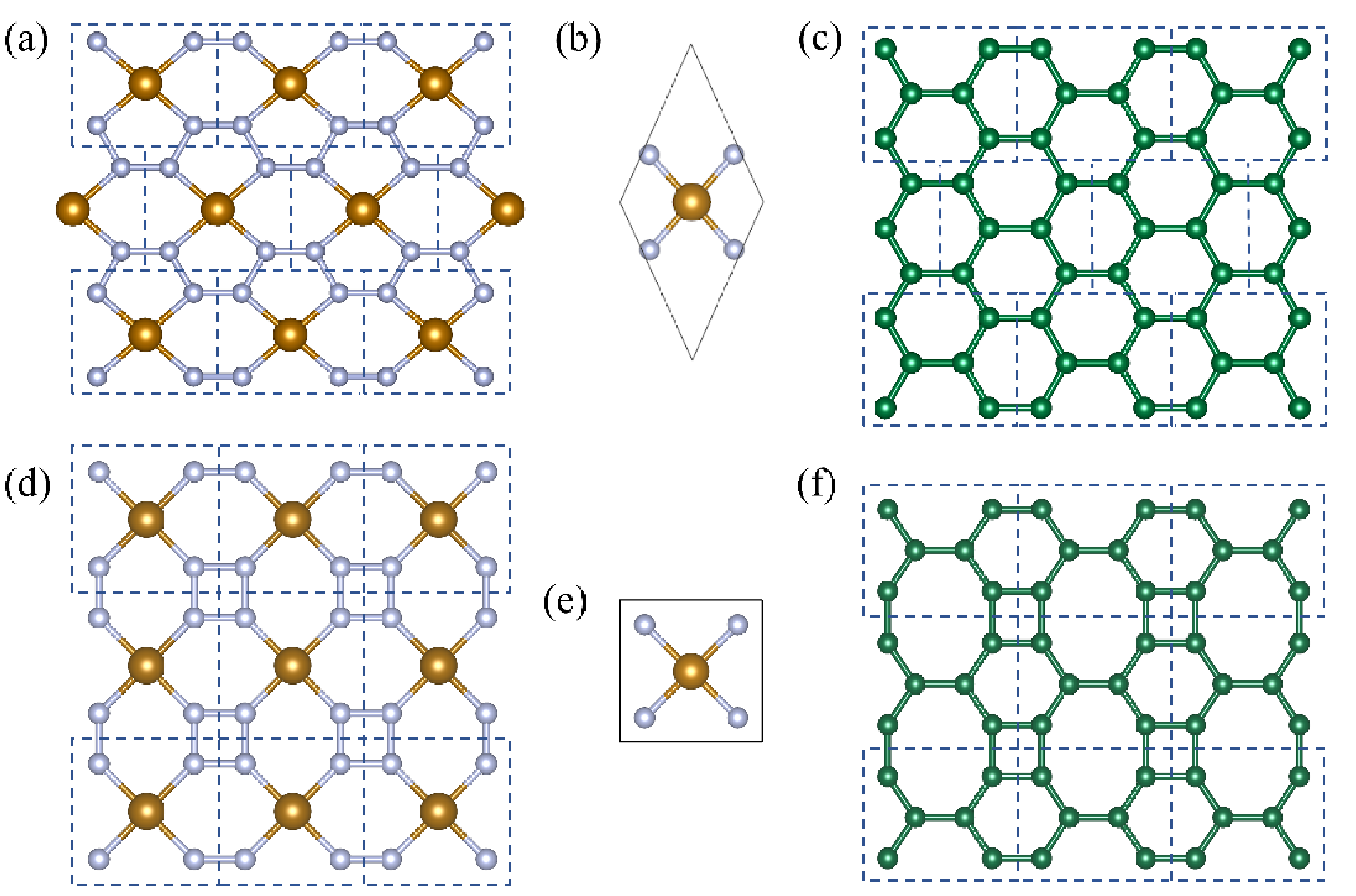}
\caption{(a) and (d), atomic structures of $r$-$M$N$_4$ and $s$-$M$N$_4$ ($M$ = V, Cr, Mn, Fe, Co) monolayers. (b) and (e), the primitive unit cells of $r$-$M$N$_4$ and $s$-$M$N$_4$. (c) and (f), the atomic structures of graphene and biphenylene network that can be divided into H-type six-carbon blocks.
 } \label{structmodel}
\end{center}
\end{figure}

We design two kinds of $M$N$_4$ ($M$ = V, Cr, Mn, Fe, Co) lattices and call them $r$-$M$N$_4$ and $s$-$M$N$_4$ in terms of the rhombic and square shapes of crystal cell. The structures are shown in Fig. \ref{structmodel}(a) and (d) and both of them are made up of $M$N$_4$ moiety marked with a dashed line rectangle. When the $M$N$_4$ moieties are linked together in the staggered and top-and-bottom alignments, the $r$-$M$N$_4$ and $s$-$M$N$_4$ are formed, which are related to the rhombic lattice with the plane group $Cmm$ symmetry and the square lattice with the plane group $P4m$ symmetry, respectively.
The primitive unit cells of two lattices are displayed in Fig. \ref{structmodel}(b) and (e), in which only one formula unit cell is included.
The structure of Graphene is shown in Fig. \ref{structmodel}(c), which is the most famous two-dimensional material. Another carbon allotrope, biphenylene network \cite{Fan2021}, had been synthesized recently, and the atomic structure is shown in Fig. \ref{structmodel}(d).
The lattices of graphene and biphenylene network can be divided into the H-type units of six carbon atoms presented by the dashed line rectangles.
By substituting $M$N$_4$ moiety for H-type unit in two carbon 2D compounds, the $r$-$M$N$_4$ and $s$-$M$N$_4$ structures are derived.
The optimized lattice parameters of $r$-$M$N$_4$ and $s$-$M$N$_4$ structures are listed in Table. \ref{latt}. We can see that from V, Cr, Mn, Fe, to Co the lattice parameters decrease because of the decrease of their atomic radius.

\begin{table}
  \caption{The optimized lattice parameters of $M$N$_4$ ($M$ = V, Cr, Mn, Fe, Co) primitive cell and the formation energy per atom are listed. The formation energy of nitrides CuN$_3$ \cite{Wilsdorf1948}, PtN$_2$ \cite{Crowhurst2006}, and $g$-C$_3$N$_4$ \cite{Groenewolt2005} are also calculated for comparison.
  }
  \label{latt}
   \begin{tabular*}{8.5cm}{lccclcc}
    \hline
     &$a$(\AA) &$\alpha$($^\circ$) &E$_{form}$(eV)& &$a$(\AA)&E$_{form}$(eV)\\
    \hline
    $r$-VN$_4$   &4.85   &134.08 & 0.40  &$s$-VN$_4$   &4.24 & 0.52   \\
    $r$-CrN$_4$  &4.78   &133.52 & 0.33  &$s$-CrN$_4$  &4.22 & 0.45   \\
    $r$-MnN$_4$  &4.61   &131.96 & 0.40  &$s$-MnN$_4$  &4.09 & 0.53   \\
    $r$-FeN$_4$  &4.51   &131.31 & 0.46  &$s$-FeN$_4$  &4.02 & 0.63   \\
    $r$-CoN$_4$  &4.44   &130.87 & 0.42  &$s$-CoN$_4$  &3.95 & 0.63   \\
    PtN$_2$   &   &   &0.42  & CuN$_3$ & &0.55\\
    $g$-C$_3$N$_4$   &  &   &0.35 &  & &\\
    \hline
  \end{tabular*}
\end{table}

\subsection{Structural Stability}

\begin{figure*}[htbp]
\begin{center}
\includegraphics[width=14cm]{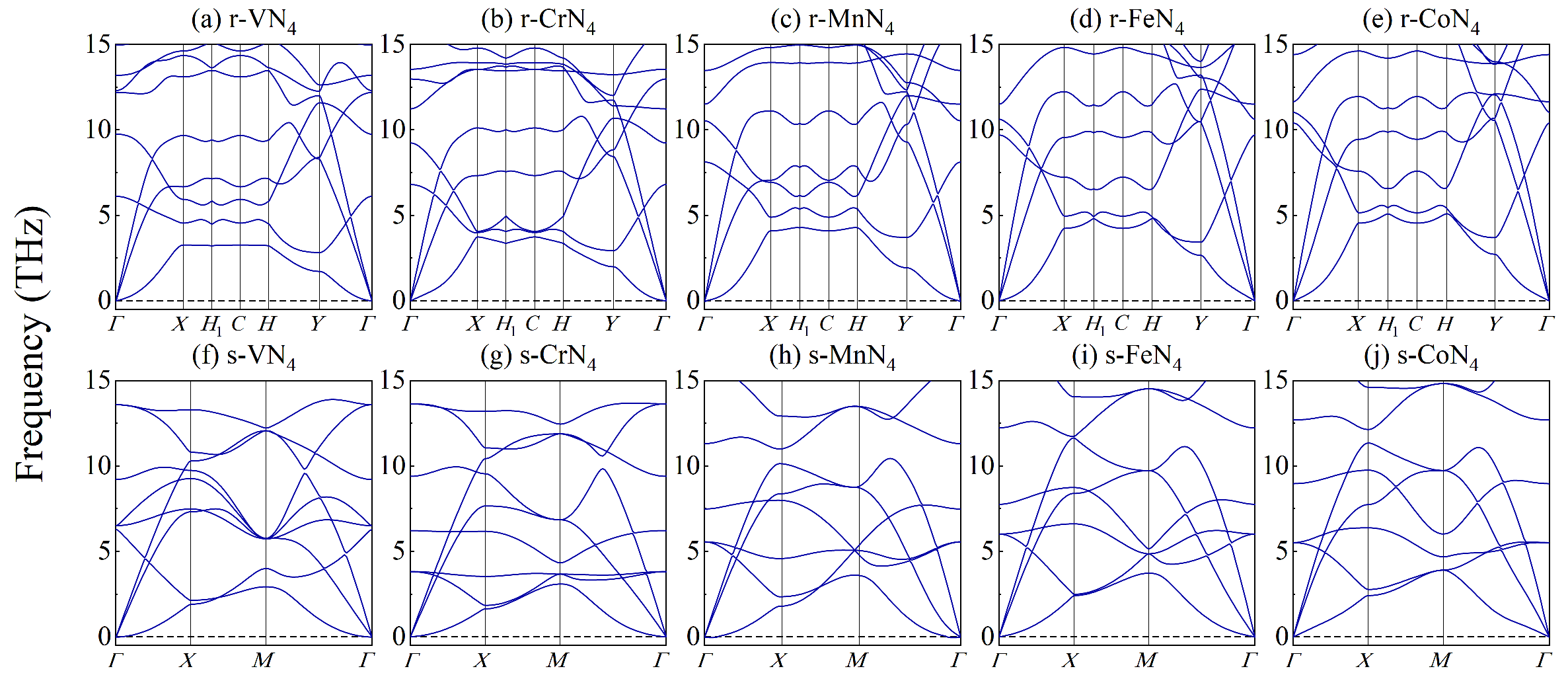}
\caption{Phonon spectra of $r$-MN$_4$ and $s$-MN$_4$ with M = V, Cr, Mn, Fe, and Co.
  } \label{phonon}
\end{center}
\end{figure*}

\begin{figure*}[htbp]
\begin{center}
\includegraphics[width=17cm]{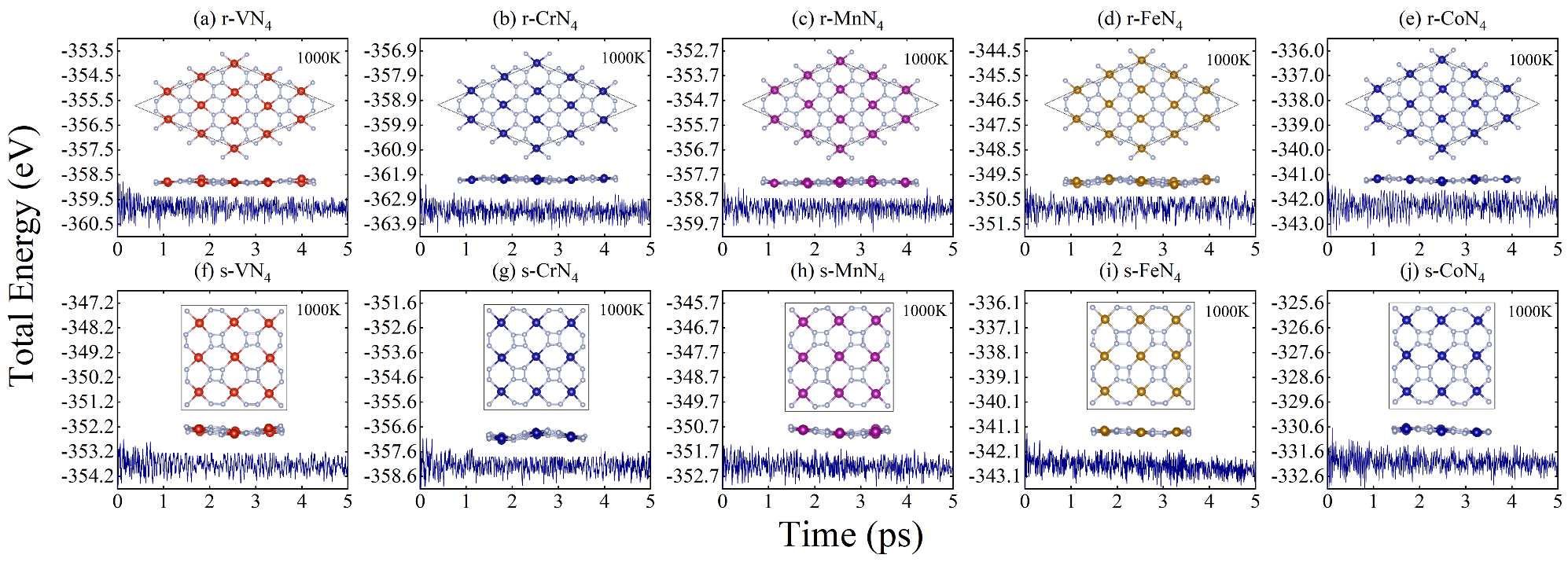}
\caption{Total energy evolutions of $r$-MN$_4$ and $s$-MN$_4$ (M = V, Cr, Mn, Fe, and Co) monolayers with respect to time in molecular dynamics simulations. The insets are the top and side views of final configurations for $r$-MN$_4$ and $s$-MN$_4$ monolayers at 1000 K after 5 ps.
  } \label{MD}
\end{center}
\end{figure*}

To inspect the structural stability of $r$-$M$N$_4$ and $s$-$M$N$_4$ ($M$ = V, Cr, Mn, Fe, and Co), we perform the formation energy, phonon spectrum, and molecular dynamics calculations.
The formation energy ${E}_{form}$ is defined as
$E_{form} = \frac{1}{n}*(E_{tot} - E_{metal} - 2 E_{N_2})$,
 in which $E_{tot}$, $E_{metal}$, and $E_{N_2}$ are the total energy, bulk metal energy per atom, and nitrogen molecule energy, respectively. The formation energies for $r$-$M$N$_4$, $s$-$M$N$_4$, and three reference compounds are listed in the Table. \ref{latt}, and the energies of $r$-$M$N$_4$ and $s$-$M$N$_4$ are comparable to the ones of nitrides CuN$_3$ \cite{Wilsdorf1948}, PtN$_2$ \cite{Crowhurst2006}, and $g$-C$_3$N$_4$ \cite{Groenewolt2005} which have already been synthesized in experiments.
Then, we perform phonon spectra calculations for $r$-$M$N$_4$, $s$-$M$N$_4$ ($M$ = V, Cr, Mn, Fe, and Co) structures using the primitive cell in ferromagnetic phase or nonmagnetic phase.
The phonon spectra of $r$-$M$N$_4$ and $s$-$M$N$_4$ ($M$ = V, Cr, Mn, Fe, and Co) are presented in Fig. \ref{phonon} (a) $\sim$ (j).
There is no imaginary frequency mode appearing in the spectra, which verifies that these structures are dynamically stable.
Finally, the thermal stability is confirmed by ab initio molecular dynamic simulations. The fluctuations of the total potential energy of the $r$-$M$N$_4$ and $s$-$M$N$_4$ monolayers and the final configurations at the end of the simulation are presented in Fig. \ref{MD} (a) $\sim$ (j). For the above structures, no distinct energy drop happens in the process of the evolution of total energy with time and the structures are well maintained after 5 ps at the temperature of 1000 K.
So, the single-atom-thick layers of $r$-$M$N$_4$ and $s$-$M$N$_4$ ($M$ = V, Cr, Mn, Fe, and Co) are energetically, dynamically, and thermally stable.

\subsection{Stability mechanism}
Next, we analyze the stability mechanism.
In $r$-$M$N$_4$ and $s$-$M$N$_4$ lattices, N atom is tricoordinated by two neighbored N atoms and one metal atom. For the five valent electrons of N atom, two of them occupy two $sp^2$ hybrid orbitals, combined with the $sp^2$ orbitals of the neighbored N atoms to form the N-N $\sigma$ bonds, while other two electrons occupy another $sp^2$ orbital which serve as lone pair electrons to form the N-metal coordination bond. These N-N $\sigma$ bonds and N-metal coordination bonds make up the framework of the planar structures. As for the remaining electron of N atom, it occupies the $p_z$ orbital perpendicular to the lattice plane. The hybridization of the N $p_z$ orbitals lead to the formation of $\pi$ bonds which result in the planarity of the square N$_4$ block and N-chain block in $s$-$M$N$_4$ and $r$-$M$N$_4$ lattices.
Besides, transition metal atom is situated at the center of four N atoms and is confined in the lattice plane, where N $p_z$ orbital is combined with the $d_{xz}/d_{yz}$ orbital of transition metal atoms to form the $\pi$ bonds. The hybridized $p_z$-$p_z$ and $p_z$-$d_{xz}/d_{yz}$ orbitals can extend over the whole monolayer which lowers the total energy of the system and further increase the strength of N-N and N-metal bonds.
Notably, it is the delocalized $\pi$ orbitals that play a decisive role in the formation of the planar structure of the two conjugated nitride monolayers.


\begin{figure}[htbp]
\begin{center}
\includegraphics[width=8.0cm]{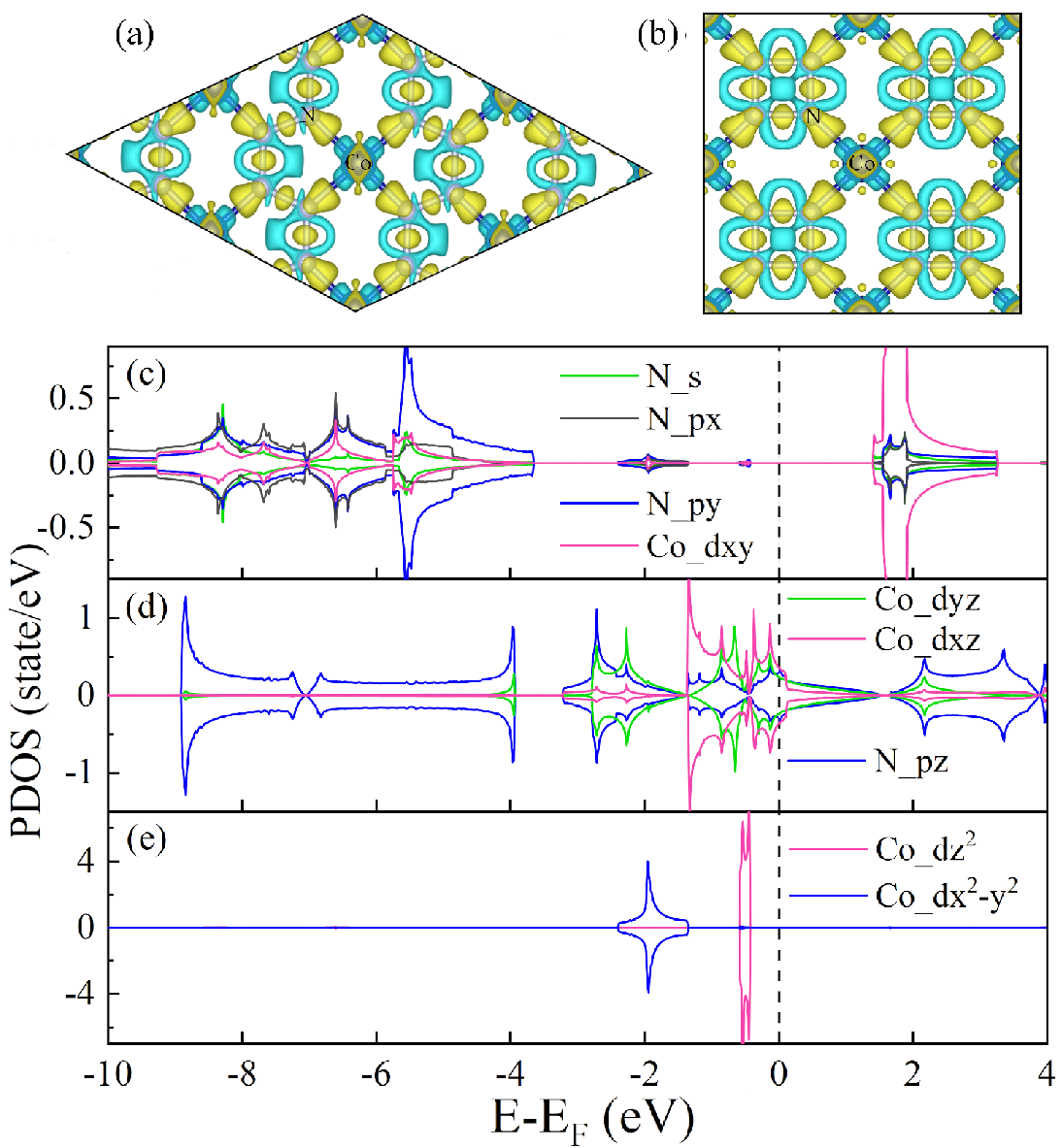}
\caption{Charge density difference relative to the superposition of atomic charge density for $r$-CoN$_4$ (a) and $s$-CoN$_4$ (b) monolayers. (c),(d),(e) Partial density of states of Co 3$d$ and N 2$p$ suborbitals in $r$-CoN$_4$ sheet.
  } \label{diff-Charg}
\end{center}
\end{figure}
The $r$-CoN$_4$ and $s$-CoN$_4$ monolayers are taken as examples to display the electronic structures of $r$-$M$N$_4$ and $s$-$M$N$_4$ ($M$ = V, Cr, Mn, Fe, and Co). Fig. \ref{diff-Charg}(a) and \ref{diff-Charg}(b) show the charge density difference of $r$-CoN$_4$ and $s$-CoN$_4$ relative to atomic charge density, in which yellow and blue surfaces are related to the charge accumulation and depletion. Yellow surfaces between N-N atoms and N-Co atoms indicate the formation of N-N covalent bond and N-Co coordination bond,
while three yellow pockets around N atom represent three $sp^2$ orbitals.
Fig. \ref{diff-Charg}(c) shows the projected density of states of $r$-CoN$_4$ on atomic orbitals. The states of $s$, $p_x$, and p$_y$ orbitals distribute in the almost same energy range, which is consistent with the $sp^2$ hybridization of $s$, $p_x$, and p$_y$ orbitals. Furthermore, the Co $d_{xy}$ orbital also has the similar distribution in energy space, which indicates one $sp^2$ orbital of N atom links to the Co $d_{xy}$ orbital to form a $\sigma$-type coordination bond.
Fig. \ref{diff-Charg}(d) shows the N $p_z$ and Co $d_{xz}$/$d_{yz}$ orbitals distributed in the same energy scope, which corresponds the formation of $\pi$ bond from the hybridization of $p_z$ and $d_{xz}$/$d_{yz}$ orbitals.
For N atom, the summary of charges in $s$, $p_x$, and p$_y$ orbitals are 3.81 electron, which is in agreement with the picture of $sp^2$ hybridization and lone pair electrons filled in one $sp^2$ orbital.
In addition, the length of N-N and Co-N bonds in $s$-$M$N$_4$ and $r$-$M$N$_4$ are 1.34 $\sim$ 1.36 \AA, which is between single N-N bond (1.45 \AA) and double N-N bond  (1.25 \AA). The length of Co-N bond is 1.82 \AA, less than the length of Co-N ionic bond (about 1.86 \AA~ or more). The shortening of N-N and Co-N bonds is ascribed to the conjugated effect, similar to the short C-C bond in graphene due to the large $\pi$ bond.

\subsection{Structural compatibility with different metals}

The $r$-$M$N$_4$ and $s$-$M$N$_4$ lattices have broad structural compatibility.
Namely, no matter which metal of V, Cr, Mn, Fe, or Co is adopted, the $r$-$M$N$_4$ and $s$-$M$N$_4$ lattices can maintain the planar structure.
The compatibility is tightly related to the N-$M$ coordination bond which is dominated by lone pair electrons of N atom.
In a planar square crystal field, $3d$ orbitals split into $d_{z^2}$, $d_{x^2-y^2}$, $d_{xz}/d_{yz}$, and $d_{xy}$ orbitals \cite{Kroll2012},
while in a rectangle field, the degenerated $d_{xz}$ and $d_{yz}$ are further split because of the unquivalence of $x$ and $y$ axes.
The $d_{xy}$ orbital of V, Cr, Mn, Fe, and Co atoms has the highest energy and is empty, which provides a favorable condition to the formation of coordination bond between N atom and V, Cr, Mn, Fe, or Co atom.
Another factor is the $\pi$-d conjugation between $d_{xz}/d_{yz}$ of transition metal and N-N $\pi$ bond belonging to N-square and N-chain blocks. The extended $\pi$ bond has a good capacity to accommodate electrons, and allows that the electron number in $d_{xz}/d_{yz}$ orbitals of metal atom is different.
The third factor is the charge transfer from $3d$ metal atom to four N atom, and the transferred charge is about one electron and only from 4$s$ electrons of metal atom.
The coordination bond, large $\pi$ bond, and the loss of 4$s$ electron is suitable for V, Cr, Mn, Fe, Co and not exclusive to a special kind of $3d$ metal.
Consequently, the $r$-$M$N$_4$ and $s$-$M$N$_4$ lattices have good comparability for different $3d$ metal atoms.
The compatibility and stability mechanism are similar to that in Ni$_3$(HITP)$_2$ and phthalocyanine compounds \cite{Chen2015}.

\subsection{Continuous adjustability of electronic and magnetic properties}
\begin{figure}[htbp]
\begin{center}
\includegraphics[width=7.0cm]{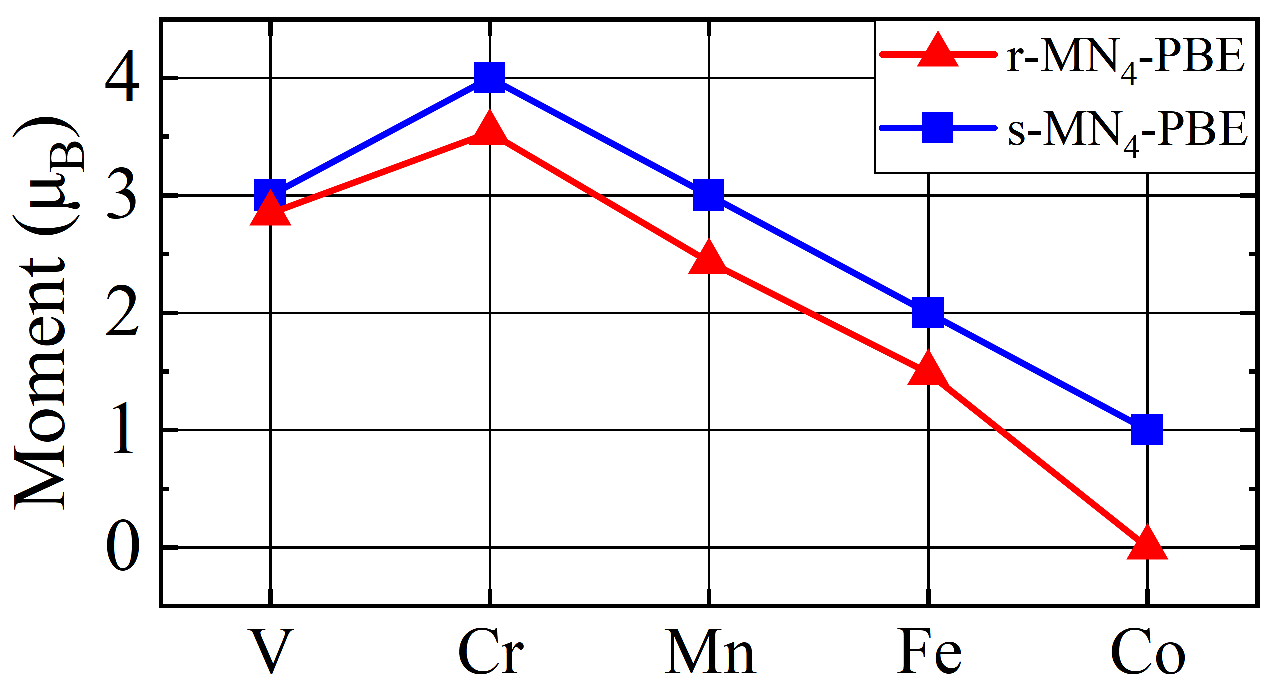}
\caption{
The variation of magnetic moments of $M$ atoms in $r$-$M$N$_4$ and $s$-$M$N$_4$ monolayers from V, Cr, Mn, Fe, to Co.
  } \label{M-variation}
\end{center}
\end{figure}

The continuous adjustability of electronic structure is a notable feature of $r$-$M$N$_4$ and $s$-$M$N$_4$ lattices. 
With transition metal atom varying from V, Cr, Mn, Fe, to Co, the total charges of $r$-$M$N$_4$ and $s$-$M$N$_4$ systems increase one electron in turn.
Meanwhile, because the charge loss of metal atom is from their $4s$ electrons and about one electron, the $d$ electrons of metal ions in $r$-$M$N$_4$ and $s$-$M$N$_4$ lattices also increase one electron gradually, resulting in the easy tunability of the electronic and magnetic properties.
In $r$-$M$N$_4$ monolayer, from V, Cr, Mn to Fe, the moment changes from 2.85 $\mu_B$, 3.54 $\mu_B$, 2.43 $\mu_B$ to 1.48 $\mu_B$ with the variation of electronic occupation in spin-up and spin-down states of $d$ orbitals.
The moment of Co atom in $r$-CoN$_4$ lattice is zero because the electronic states are unpolarized.
For the $s$-$M$N$_4$ lattice, the moments of V, Cr, Mn, Fe, and Co atoms are 3.0 $\mu_B$, 4.0 $\mu_B$, 3.0 $\mu_B$, 2.0 $\mu_B$, and 1.0 $\mu_B$, respectively. The magnitude of these moments is an integer multiple of Bohr magneton, which is very rare in previous studies.
The variation of magnetic moment in $r$-MN$_4$ and $s$-MN$_4$ monolayer from V, Cr, Mn, Fe to Co is shown in Fig. \ref{M-variation}.

\begin{figure*}[htbp]
\begin{center}
\includegraphics[width=14.0cm]{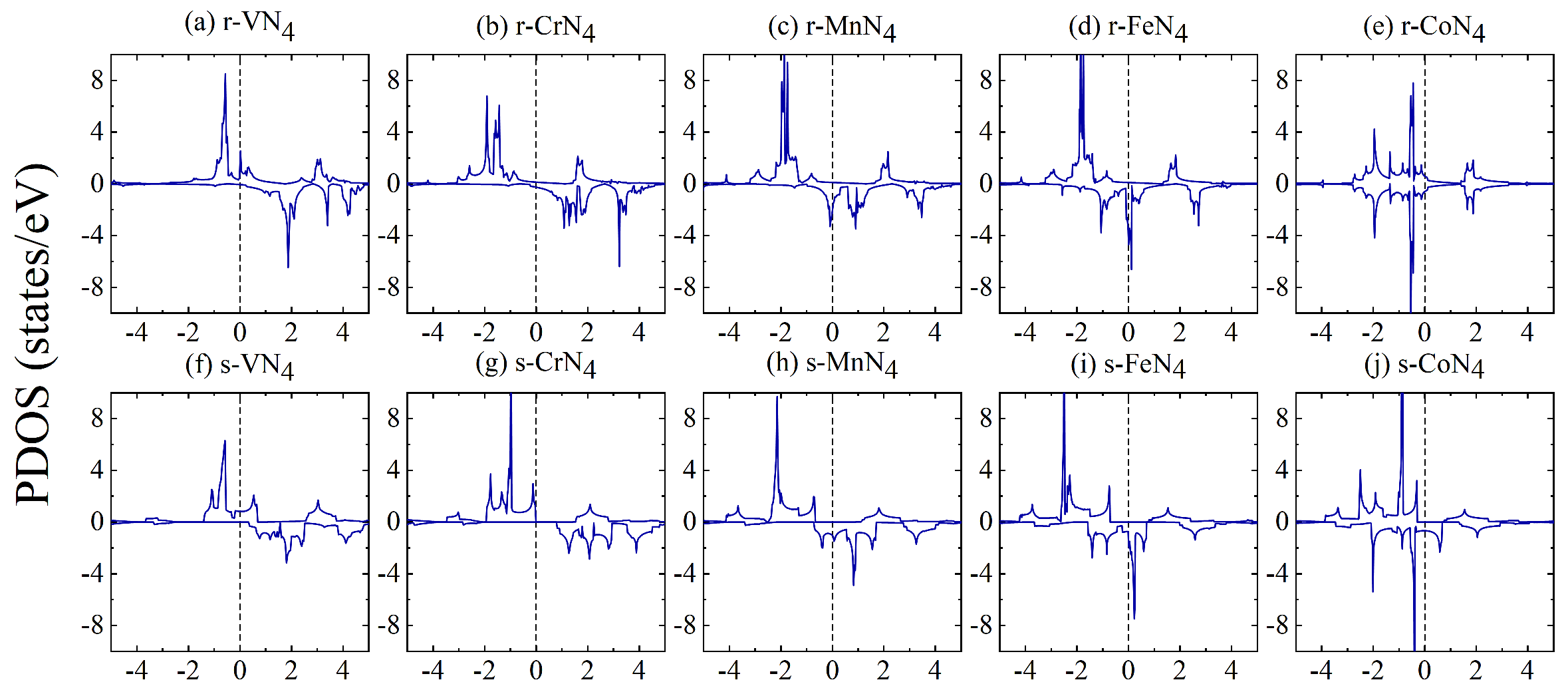}
\caption{Projected DOS on d orbitals of $M$ atom in $r$-$M$N$_4$ and $s$-$M$N$_4$ monolayers with M = V, Cr, Mn, Fe, or Co. The Fermi energy is set to zero. From V, Cr, Mn, Fe to Co, more and more $d$ states are occupied.
  } \label{PBE-d-occpu}
\end{center}
\end{figure*}
Fig. \ref{PBE-d-occpu} shows the projected DOS on $d$ orbitals of $M$ atoms in $r$-$M$N$_4$ and $s$-$M$N$_4$ monolayers with $M$ = V, Cr, Mn, Fe, and Co. From V to Cr in $r$-$M$N$_4$ monolayer, more spin-up states of $d$ orbitals are filled while spin-down states still keep empty, which leads to the magnetic moment increase from 2.85 $\mu_B$ to 3.54 $\mu_B$ . From Cr, Mn to Fe, the electron population in spin-up states of $d$ oribtals remain unchanged, but more and more spin-down states are occupied by electrons, which results in the moment decrease from 3.54 $\mu_B$, 2.43 $\mu_B$ to 1.48 $\mu_B$. The $r$-CoN$_4$ monolayer is nonmagnetic, the spin-up and spin-down states of Co $d$ orbitals are degenerate and the moment of Co atom is zero.  For V, Cr, Mn, Fe, and Co atoms in the $s$-$M$N$_4$ monolayers, the electron occupations of $d$ orbitals are similar to the scenarios in $r$-$M$N$_4$ monolayers. The gradual change in the number of electrons occupied in the $d$ orbitals demonstrates the uniqueness of these $r$-$M$N$_4$ and $s$-$M$N$_4$ monolayers.

This continuous adjustability of electronic structures is tightly related to the planar crystal field, which is another feature of $r$-$M$N$_4$ and $s$-$M$N$_4$ lattices.
The planar crystal field in $s$-$M$N$_4$ or $r$-$M$N$_4$ has a square or rectangle symmetry, which drives five suborbitals of 3$d$ metal to split in a special mode and then bring about distinctive electronic and magnetic phenomenons.
For 3$d$ metal atom in $r$-$M$N$_4$ and $s$-$M$N$_4$, because of no atom in their extension directions of d$_{z^2}$ and d$_{x^2-y^2}$ orbitals, there is no or little hybridization between the two suborbitals and N 2$p$ orbitals. In particular, the d$_{z^2}$ orbital is an almost isolated orbital.
We take $r$-CoN$_4$ as an example to address this. In Fig. \ref{diff-Charg}(e), the energy ranges of Co d$_{z^2}$ and d$_{x^2-y^2}$ states have small overlap to the range of N 2$p$ states, and the density of state of d$_{z^2}$ orbital is localized in a narrow energy scope and takes on a sharp peak.

\subsection{Ferromagnetism in the $s$-CoN$_4$ monolayer}

\begin{table}
\caption{The energies of $r$-$M$N$_4$ and $s$-$M$N$_4$ in FM, AFM-I, and AFM-II orders from the PBE calculations. The FM energy is set to zero. The units of energy is meV. The dashed line means that the initial magnetic phase would evolve into the nonmagnetic phase in the spin-polarized calculations.
	}
	\label{energy-pbe}
\begin{tabular}{lccc}
	 \hline
	PBE & E$_{FM}$ & E$_{AFM-I}$ & E$_{AFM-II}$  \\
	 \hline
	              r-VN4	    &	0	&	-97.50	&	-11.06	\\
                  r-CrN4	&	0	&	52.02	&	-68.56	\\
                  r-MnN4	&	0	&	90.95	&	-28.24	\\
                  r-FeN4	&	0	&	17.36   &	1.01	\\
                  r-CoN4	    &	-- --	&	-- --	&	-- --	\\
                  s-VN4	    &	0	&	57.04	&	-52.28	\\
                  s-CrN4	&	0	&	-128.15 &	-34.67	\\
                  s-MnN4	&	0	&	-81.54	&	54.58	\\
                  s-FeN4	&	0	&	-25.43	&	27.42	\\
                    s-CoN4	&	0	&	-- --	&	18.56	\\
                    	 \hline
\end{tabular}
\end{table}

\begin{table}
\caption{The energies of $r$-$M$N$_4$ and $s$-$M$N$_4$ in FM, AFM-I, and AFM-II orders from the SCAN calculations. The FM energy is set to zero. The units of energy is meV.
	}
	\label{energy-scan}
\begin{tabular}{lccc}
	 \hline
	SCAN & E$_{FM}$ & E$_{AFM-I}$ & E$_{AFM-II}$  \\
	 \hline
                  r-VN4	 &	0	&	-118.98	&	-33.46  \\
                  r-CrN4	&	0	&	45.35	&	-67.41 \\
                  r-MnN4	&	0	&	107.09    &	-61.99 \\
                   r-FeN4	&	0	&	36.59	&	-50.43 \\
                  r-CoN4	&	0	&	-88.48	&	-28.13 \\
                  s-VN4	&	0	&	76.10	&	-47.49 \\
                  s-CrN4	&	0	&	-94.78	&	-20.43 \\
                  s-MnN4	&	0	&	-95.03	&	10.82 \\
                    s-FeN4	&	0	&	44.76	&	 23.02 \\
                   s-CoN4	&	0	&	122.93	&	28.35 \\
                    	 \hline
\end{tabular}
\end{table}

\begin{figure*}[htbp]
\begin{center}
\includegraphics[width=14.0cm]{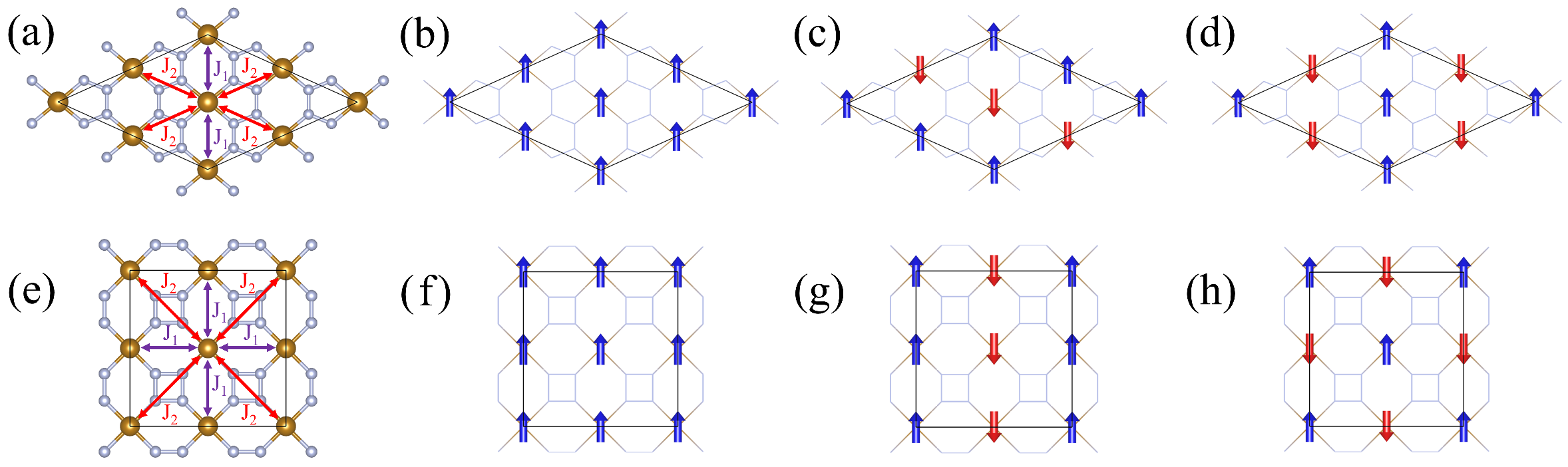}
\caption{(a) and (e), the 2 $\times$ 2 supercell of $r$-$M$N$_4$ and $s$-$M$N$_4$ lattices and the exchange interactions (J$_1$ and J$_2$) between two neighbored magnetic atoms. (b) - (d) and (f) - (h), the sketches of FM, AFM-I, and AFM-II orders for $r$-$M$N$_4$ and $s$-$M$N$_4$ lattices, in which the red and blue arrows represent two kinds of magnetic moments in opposite directions.
  } \label{orders}
\end{center}
\end{figure*}

\begin{figure}[htbp]
\begin{center}
\includegraphics[width=7.0cm]{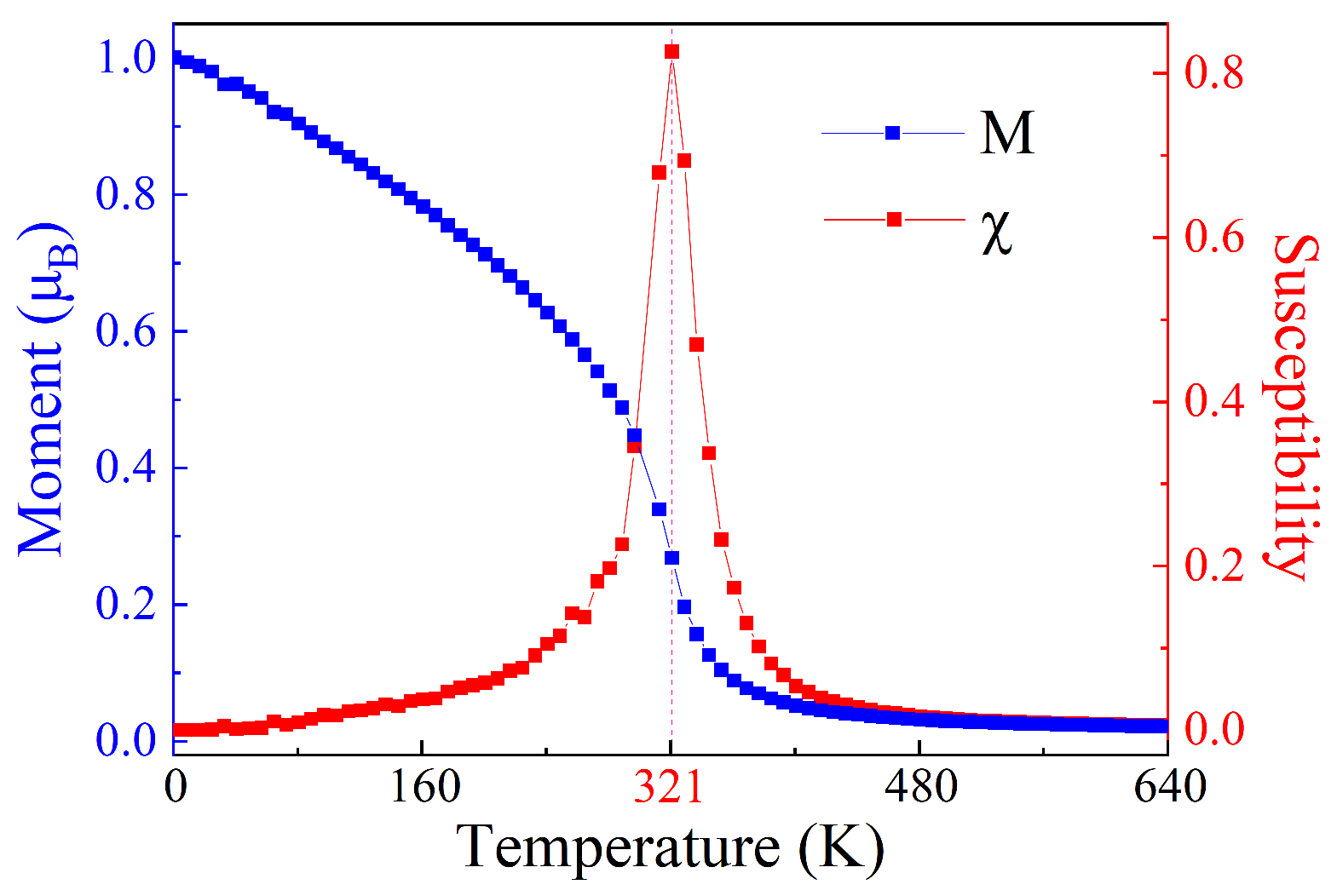}
\caption{The variation of the magnetic moment ($M$) and susceptibility ($\chi$ = $\frac{<\vec{M}^2> - <\vec{M}>^2}{k_BT}$) with respect to temperature in the $s$-CoN$_4$ monolayer.
  } \label{Tc}
\end{center}
\end{figure}

In 2017, the 2D ferromagnetism was first discovered experimentally in CrI$_3$ monolayer \cite{Huang2017}. Since then, the researches on the ferromagnetic 2D materials have attracted great interest due to their potential application in the next-generation spintronic devices including giant magneto-resistance, spin valve, magnetic random-access memory and other spin logic devices \cite{Burch2018, Shabbir2018,Sethulakshmi2019,Cortie2020}.
In the following content, we compute the energies of $r$-$M$N$_4$ and $s$-$M$N$_4$ monolayers in the several magnetic orders to screen out the ferromagnetic $s$-CoN$_4$ monolayer.
Fig. \ref{orders} (a) and (e) display the 2 $\times$ 2 supercells of $r$-$M$N$_4$ and $s$-$M$N$_4$ monolayers, respectively.  The sketches of ferromagnetic (FM) order and two antiferromagnetic (AFM) orders (AFM-I and AFM-II) in $r$-$M$N$_4$ and $s$-$M$N$_4$ lattices are displayed separately in Fig. \ref{orders} (b) $\sim$ (d) and (f) $\sim$ (h), in which the red and blue arrows represent two kinds of magnetic moments in opposite directions.
To clarify the magnetism in the ground state of $r$-$M$N$_4$ and $s$-$M$N$_4$ systems, we compute the energies of the three magnetic orders with the PBE functional method and SCAN meta-GGA functional method, and the data are listed in Table. \ref{energy-pbe} and Table. \ref{energy-scan}.
The PBE calculations manifest that $r$-FeN$_4$ and $s$-CoN$_4$ are ferromagnetic in the ground states and $r$-CoN$_4$ is nonmagnetic, while others are antiferromagnetic.
The results from the SCAN calculations are slightly different from the PBE results, which indicates that $s$-FeN$_4$ and $s$-CoN$_4$ are ferromagnetic and the others are antiferromagnetic.
Considering the results from the PBE and SCAN calculations together, we screen out the $s$-CoN$_4$ monolayer to investigate its ferromagnetism.

Curie temperature is a crucial parameter of ferromagnetic 2D materials, which determines the application prospects in electronic devices.
Next, we evaluate the Curie temperature of the $s$-CoN$_4$ monolayer with the following method.
 Firstly, as is the case with many previous studies \cite{Ma2009,Yu2015,Zhao2009},
Heisenberg model is adopted to describe the magnetic interactions in the $s$-CoN$_4$ monolayer, and the exchange couplings are obtained from the first principles calculations. Then, the Hamiltonian is solved by the Monte Carlo method and the Curie temperature is determined.
The expressions of Heisenberg model is \cite{Zhang2021}
\begin{equation}
\label{Heisenberg}
H = J_{1}\sum_{<ij>}\vec{{S}_{i}}\cdot \vec{{S}_{j}} + J_{2}\sum_{\ll ij^{\prime} \gg}\vec{{S}_{i}}\cdot \vec{{S}_{j^{\prime}}} + A\sum_{i}(S_{iz})^2,
\end{equation}
where $j$ and $j^\prime$ denote the nearest and next-nearest neighbors of $i$ site, and
$J_1$ and $J_2$ are the nearest and next-nearest neighbor couplings.
Because of the good performance of the SCAN meta-GGA method in simulating the electronic structure of 3$d$ metal compounds \cite{Sun2015}, the energies from the SCAN calculations are adopted.
In terms of the energies of $s$-CoN$_4$ in Table. \ref{energy-scan}, the nearest neighbor coupling $J_1$ and next-nearest neighbor coupling $J_2$ in the 2D magnetic lattice can be derived \cite{Feng2022}, which are -7.09 meV/S$^2$ and -27.19 meV/S$^2$, respectively. The coupling over longer distance $J_3$ is -0.73 meV/S$^2$ when a ferrimagnetic order is taken into account (not shown).
For ferromagnetic $s$-CoN$_4$ monolayer, the Curie temperature is determined to be 321 K.
The spontaneous magnetization ($M$) and susceptibility ($\chi$) of $s$-CoN$_4$ monolayer with respect to temperature are plotted in Fig. \ref{Tc}.
So, the $s$-CoN$_4$ monolayer is a room-temperature ferromagnetic two-dimensional material.

\section{Conclusion}

In summary, by the first-principles calculations, we predict two kinds of $M$N$_4$ ($M$ =metal) two-dimensional lattices, which are single-atom-thick transition metal nitride sheets with a planar structure.
The formation energy and phonon dispersion calculations as well as molecular dynamics simulations are carried out, and the $r$-$M$N$_4$ and $s$-$M$N$_4$ ($M$ = V, Cr, Mn, Fe, and Co) sheets are demonstrated to be dynamically and thermally stable.
The mechanism of structural stability is ascribed to the $sp^2$ hybridization and coordination bond between N and metal atoms.
Apart from this, the conjugation effect of $\pi$-$\pi$ interaction and the $\pi$-d coupling further stabilize the two lattices, and the extended $\pi$ bond plays a key role in keeping a planar configuration.
On the other hand, the two lattices have a good structural compatibility for different transition metals, resulting in the easy tunability and designability of electronic structure.
The planar ligand field is another feature of the two lattices, which brings about the special splitting of five suborbitals of $3d$ metal atom and gives rise to large local moment and strong magnetism.
The $s$-CoN$_4$ monolayer is a room-temperature ferromagnetic two-dimensional material and the Curie temperature of 321 K is evaluated by solving the Heisenberg Hamiltonian with the Monte Carlo method.

This work was supported by the National Natural Science Foundation of China (Grants Nos. 11974207, 11974194, 11774420, 11474004, 11704006), the National R\&D Program of China (Grants No. 2017YFA0302900), and the Major Basic Program of Natural Science Foundation of Shandong Province (Grant No. ZR202105280001).


\bibliography{Ref}

\begin{thebibliography}{38}%
\makeatletter
\providecommand \@ifxundefined [1]{%
 \@ifx{#1\undefined}
}%
\providecommand \@ifnum [1]{%
 \ifnum #1\expandafter \@firstoftwo
 \else \expandafter \@secondoftwo
 \fi
}%
\providecommand \@ifx [1]{%
 \ifx #1\expandafter \@firstoftwo
 \else \expandafter \@secondoftwo
 \fi
}%
\providecommand \natexlab [1]{#1}%
\providecommand \enquote  [1]{``#1''}%
\providecommand \bibnamefont  [1]{#1}%
\providecommand \bibfnamefont [1]{#1}%
\providecommand \citenamefont [1]{#1}%
\providecommand \href@noop [0]{\@secondoftwo}%
\providecommand \href [0]{\begingroup \@sanitize@url \@href}%
\providecommand \@href[1]{\@@startlink{#1}\@@href}%
\providecommand \@@href[1]{\endgroup#1\@@endlink}%
\providecommand \@sanitize@url [0]{\catcode `\\12\catcode `\$12\catcode
  `\&12\catcode `\#12\catcode `\^12\catcode `\_12\catcode `\%12\relax}%
\providecommand \@@startlink[1]{}%
\providecommand \@@endlink[0]{}%
\providecommand \url  [0]{\begingroup\@sanitize@url \@url }%
\providecommand \@url [1]{\endgroup\@href {#1}{\urlprefix }}%
\providecommand \urlprefix  [0]{URL }%
\providecommand \Eprint [0]{\href }%
\providecommand \doibase [0]{http://dx.doi.org/}%
\providecommand \selectlanguage [0]{\@gobble}%
\providecommand \bibinfo  [0]{\@secondoftwo}%
\providecommand \bibfield  [0]{\@secondoftwo}%
\providecommand \translation [1]{[#1]}%
\providecommand \BibitemOpen [0]{}%
\providecommand \bibitemStop [0]{}%
\providecommand \bibitemNoStop [0]{.\EOS\space}%
\providecommand \EOS [0]{\spacefactor3000\relax}%
\providecommand \BibitemShut  [1]{\csname bibitem#1\endcsname}%
\let\auto@bib@innerbib\@empty
\bibitem [{\citenamefont {Salamat}\ \emph {et~al.}(2013)\citenamefont
  {Salamat}, \citenamefont {Hector}, \citenamefont {Kroll},\ and\ \citenamefont
  {McMillan}}]{Salamat2013}%
  \BibitemOpen
  \bibfield  {author} {\bibinfo {author} {\bibfnamefont {A.}~\bibnamefont
  {Salamat}}, \bibinfo {author} {\bibfnamefont {A.~L.}\ \bibnamefont {Hector}},
  \bibinfo {author} {\bibfnamefont {P.}~\bibnamefont {Kroll}}, \ and\ \bibinfo
  {author} {\bibfnamefont {P.~F.}\ \bibnamefont {McMillan}},\ }\href {\doibase
  10.1016/j.ccr.2013.01.010} {\bibfield  {journal} {\bibinfo  {journal}
  {Coordination Chemistry Reviews}\ }\textbf {\bibinfo {volume} {257}},\
  \bibinfo {pages} {2063} (\bibinfo {year} {2013})}\BibitemShut {NoStop}%
\bibitem [{\citenamefont {Zheng}\ \emph {et~al.}(2020)\citenamefont {Zheng},
  \citenamefont {Li}, \citenamefont {Pi}, \citenamefont {Song}, \citenamefont
  {Gao}, \citenamefont {Chu},\ and\ \citenamefont {Huo}}]{Zheng2020}%
  \BibitemOpen
  \bibfield  {author} {\bibinfo {author} {\bibfnamefont {Y.}~\bibnamefont
  {Zheng}}, \bibinfo {author} {\bibfnamefont {X.}~\bibnamefont {Li}}, \bibinfo
  {author} {\bibfnamefont {C.}~\bibnamefont {Pi}}, \bibinfo {author}
  {\bibfnamefont {H.}~\bibnamefont {Song}}, \bibinfo {author} {\bibfnamefont
  {B.}~\bibnamefont {Gao}}, \bibinfo {author} {\bibfnamefont {P.~K.}\
  \bibnamefont {Chu}}, \ and\ \bibinfo {author} {\bibfnamefont
  {K.}~\bibnamefont {Huo}},\ }\href {\doibase 10.1016/j.flatc.2019.100149}
  {\bibfield  {journal} {\bibinfo  {journal} {FlatChem}\ }\textbf {\bibinfo
  {volume} {19}},\ \bibinfo {pages} {100149} (\bibinfo {year}
  {2020})}\BibitemShut {NoStop}%
\bibitem [{\citenamefont {Shabbir}\ \emph {et~al.}(2018)\citenamefont
  {Shabbir}, \citenamefont {Nadeem}, \citenamefont {Dai}, \citenamefont
  {Fuhrer}, \citenamefont {Xue}, \citenamefont {Wang},\ and\ \citenamefont
  {Bao}}]{Shabbir2018}%
  \BibitemOpen
  \bibfield  {author} {\bibinfo {author} {\bibfnamefont {B.}~\bibnamefont
  {Shabbir}}, \bibinfo {author} {\bibfnamefont {M.}~\bibnamefont {Nadeem}},
  \bibinfo {author} {\bibfnamefont {Z.}~\bibnamefont {Dai}}, \bibinfo {author}
  {\bibfnamefont {M.~S.}\ \bibnamefont {Fuhrer}}, \bibinfo {author}
  {\bibfnamefont {Q.~K.}\ \bibnamefont {Xue}}, \bibinfo {author} {\bibfnamefont
  {X.}~\bibnamefont {Wang}}, \ and\ \bibinfo {author} {\bibfnamefont
  {Q.}~\bibnamefont {Bao}},\ }\href {\doibase 10.1063/1.5040694} {\bibfield
  {journal} {\bibinfo  {journal} {Applied Physics Reviews}\ }\textbf {\bibinfo
  {volume} {5}} (\bibinfo {year} {2018}),\ 10.1063/1.5040694}\BibitemShut
  {NoStop}%
\bibitem [{\citenamefont {Sethulakshmi}\ \emph {et~al.}(2019)\citenamefont
  {Sethulakshmi}, \citenamefont {Mishra}, \citenamefont {Ajayan}, \citenamefont
  {Kawazoe}, \citenamefont {Roy}, \citenamefont {Singh},\ and\ \citenamefont
  {Tiwary}}]{Sethulakshmi2019}%
  \BibitemOpen
  \bibfield  {author} {\bibinfo {author} {\bibfnamefont {N.}~\bibnamefont
  {Sethulakshmi}}, \bibinfo {author} {\bibfnamefont {A.}~\bibnamefont
  {Mishra}}, \bibinfo {author} {\bibfnamefont {P.~M.}\ \bibnamefont {Ajayan}},
  \bibinfo {author} {\bibfnamefont {Y.}~\bibnamefont {Kawazoe}}, \bibinfo
  {author} {\bibfnamefont {A.~K.}\ \bibnamefont {Roy}}, \bibinfo {author}
  {\bibfnamefont {A.~K.}\ \bibnamefont {Singh}}, \ and\ \bibinfo {author}
  {\bibfnamefont {C.~S.}\ \bibnamefont {Tiwary}},\ }\href {\doibase
  10.1016/j.mattod.2019.03.015} {\bibfield  {journal} {\bibinfo  {journal}
  {Materials Today}\ }\textbf {\bibinfo {volume} {27}},\ \bibinfo {pages} {107}
  (\bibinfo {year} {2019})}\BibitemShut {NoStop}%
\bibitem [{\citenamefont {Fei}\ \emph {et~al.}(2018)\citenamefont {Fei},
  \citenamefont {Dong}, \citenamefont {Feng}, \citenamefont {Allen},
  \citenamefont {Wan}, \citenamefont {Volosskiy}, \citenamefont {Li},
  \citenamefont {Zhao}, \citenamefont {Wang}, \citenamefont {Sun},
  \citenamefont {An}, \citenamefont {Chen}, \citenamefont {Guo}, \citenamefont
  {Lee}, \citenamefont {Chen}, \citenamefont {Shakir}, \citenamefont {Liu},
  \citenamefont {Hu}, \citenamefont {Li}, \citenamefont {Kirkland},
  \citenamefont {Duan},\ and\ \citenamefont {Huang}}]{Fei2018}%
  \BibitemOpen
  \bibfield  {author} {\bibinfo {author} {\bibfnamefont {H.}~\bibnamefont
  {Fei}}, \bibinfo {author} {\bibfnamefont {J.}~\bibnamefont {Dong}}, \bibinfo
  {author} {\bibfnamefont {Y.}~\bibnamefont {Feng}}, \bibinfo {author}
  {\bibfnamefont {C.~S.}\ \bibnamefont {Allen}}, \bibinfo {author}
  {\bibfnamefont {C.}~\bibnamefont {Wan}}, \bibinfo {author} {\bibfnamefont
  {B.}~\bibnamefont {Volosskiy}}, \bibinfo {author} {\bibfnamefont
  {M.}~\bibnamefont {Li}}, \bibinfo {author} {\bibfnamefont {Z.}~\bibnamefont
  {Zhao}}, \bibinfo {author} {\bibfnamefont {Y.}~\bibnamefont {Wang}}, \bibinfo
  {author} {\bibfnamefont {H.}~\bibnamefont {Sun}}, \bibinfo {author}
  {\bibfnamefont {P.}~\bibnamefont {An}}, \bibinfo {author} {\bibfnamefont
  {W.}~\bibnamefont {Chen}}, \bibinfo {author} {\bibfnamefont {Z.}~\bibnamefont
  {Guo}}, \bibinfo {author} {\bibfnamefont {C.}~\bibnamefont {Lee}}, \bibinfo
  {author} {\bibfnamefont {D.}~\bibnamefont {Chen}}, \bibinfo {author}
  {\bibfnamefont {I.}~\bibnamefont {Shakir}}, \bibinfo {author} {\bibfnamefont
  {M.}~\bibnamefont {Liu}}, \bibinfo {author} {\bibfnamefont {T.}~\bibnamefont
  {Hu}}, \bibinfo {author} {\bibfnamefont {Y.}~\bibnamefont {Li}}, \bibinfo
  {author} {\bibfnamefont {A.~I.}\ \bibnamefont {Kirkland}}, \bibinfo {author}
  {\bibfnamefont {X.}~\bibnamefont {Duan}}, \ and\ \bibinfo {author}
  {\bibfnamefont {Y.}~\bibnamefont {Huang}},\ }\href {\doibase
  10.1038/s41929-017-0008-y} {\bibfield  {journal} {\bibinfo  {journal} {Nature
  Catalysis}\ }\textbf {\bibinfo {volume} {1}},\ \bibinfo {pages} {63}
  (\bibinfo {year} {2018})}\BibitemShut {NoStop}%
\bibitem [{\citenamefont {He}\ \emph {et~al.}(2019)\citenamefont {He},
  \citenamefont {He}, \citenamefont {Deng}, \citenamefont {Peng}, \citenamefont
  {Chen}, \citenamefont {Zhang}, \citenamefont {Yao}, \citenamefont {Zhang},
  \citenamefont {Xiao}, \citenamefont {Ma}, \citenamefont {Ge},\ and\
  \citenamefont {Ji}}]{He2019}%
  \BibitemOpen
  \bibfield  {author} {\bibinfo {author} {\bibfnamefont {X.}~\bibnamefont
  {He}}, \bibinfo {author} {\bibfnamefont {Q.}~\bibnamefont {He}}, \bibinfo
  {author} {\bibfnamefont {Y.}~\bibnamefont {Deng}}, \bibinfo {author}
  {\bibfnamefont {M.}~\bibnamefont {Peng}}, \bibinfo {author} {\bibfnamefont
  {H.}~\bibnamefont {Chen}}, \bibinfo {author} {\bibfnamefont {Y.}~\bibnamefont
  {Zhang}}, \bibinfo {author} {\bibfnamefont {S.}~\bibnamefont {Yao}}, \bibinfo
  {author} {\bibfnamefont {M.}~\bibnamefont {Zhang}}, \bibinfo {author}
  {\bibfnamefont {D.}~\bibnamefont {Xiao}}, \bibinfo {author} {\bibfnamefont
  {D.}~\bibnamefont {Ma}}, \bibinfo {author} {\bibfnamefont {B.}~\bibnamefont
  {Ge}}, \ and\ \bibinfo {author} {\bibfnamefont {H.}~\bibnamefont {Ji}},\
  }\href {\doibase 10.1038/s41467-019-11619-6} {\bibfield  {journal} {\bibinfo
  {journal} {Nature Communications}\ }\textbf {\bibinfo {volume} {10}},\
  \bibinfo {pages} {3663} (\bibinfo {year} {2019})}\BibitemShut {NoStop}%
\bibitem [{\citenamefont {Zhao}\ \emph {et~al.}(2019)\citenamefont {Zhao},
  \citenamefont {Zhang}, \citenamefont {Huang}, \citenamefont {Liu},
  \citenamefont {Zhang}, \citenamefont {He}, \citenamefont {Wu}, \citenamefont
  {Zhang}, \citenamefont {Wu}, \citenamefont {Yang}, \citenamefont {Gu},
  \citenamefont {Hu},\ and\ \citenamefont {Wan}}]{Zhao2019}%
  \BibitemOpen
  \bibfield  {author} {\bibinfo {author} {\bibfnamefont {L.}~\bibnamefont
  {Zhao}}, \bibinfo {author} {\bibfnamefont {Y.}~\bibnamefont {Zhang}},
  \bibinfo {author} {\bibfnamefont {L.-B.}\ \bibnamefont {Huang}}, \bibinfo
  {author} {\bibfnamefont {X.-Z.}\ \bibnamefont {Liu}}, \bibinfo {author}
  {\bibfnamefont {Q.-H.}\ \bibnamefont {Zhang}}, \bibinfo {author}
  {\bibfnamefont {C.}~\bibnamefont {He}}, \bibinfo {author} {\bibfnamefont
  {Z.-Y.}\ \bibnamefont {Wu}}, \bibinfo {author} {\bibfnamefont {L.-J.}\
  \bibnamefont {Zhang}}, \bibinfo {author} {\bibfnamefont {J.}~\bibnamefont
  {Wu}}, \bibinfo {author} {\bibfnamefont {W.}~\bibnamefont {Yang}}, \bibinfo
  {author} {\bibfnamefont {L.}~\bibnamefont {Gu}}, \bibinfo {author}
  {\bibfnamefont {J.-S.}\ \bibnamefont {Hu}}, \ and\ \bibinfo {author}
  {\bibfnamefont {L.-J.}\ \bibnamefont {Wan}},\ }\href {\doibase
  10.1038/s41467-019-09290-y} {\bibfield  {journal} {\bibinfo  {journal}
  {Nature Communications}\ }\textbf {\bibinfo {volume} {10}},\ \bibinfo {pages}
  {1278} (\bibinfo {year} {2019})}\BibitemShut {NoStop}%
\bibitem [{\citenamefont {Hu}\ \emph {et~al.}(2021)\citenamefont {Hu},
  \citenamefont {Wang}, \citenamefont {Tan}, \citenamefont {Duan},
  \citenamefont {Li}, \citenamefont {Li}, \citenamefont {Ji}, \citenamefont
  {Lu}, \citenamefont {Wang}, \citenamefont {Sun}, \citenamefont {Hu},\ and\
  \citenamefont {Yan}}]{Hu2021}%
  \BibitemOpen
  \bibfield  {author} {\bibinfo {author} {\bibfnamefont {W.}~\bibnamefont
  {Hu}}, \bibinfo {author} {\bibfnamefont {C.}~\bibnamefont {Wang}}, \bibinfo
  {author} {\bibfnamefont {H.}~\bibnamefont {Tan}}, \bibinfo {author}
  {\bibfnamefont {H.}~\bibnamefont {Duan}}, \bibinfo {author} {\bibfnamefont
  {G.}~\bibnamefont {Li}}, \bibinfo {author} {\bibfnamefont {N.}~\bibnamefont
  {Li}}, \bibinfo {author} {\bibfnamefont {Q.}~\bibnamefont {Ji}}, \bibinfo
  {author} {\bibfnamefont {Y.}~\bibnamefont {Lu}}, \bibinfo {author}
  {\bibfnamefont {Y.}~\bibnamefont {Wang}}, \bibinfo {author} {\bibfnamefont
  {Z.}~\bibnamefont {Sun}}, \bibinfo {author} {\bibfnamefont {F.}~\bibnamefont
  {Hu}}, \ and\ \bibinfo {author} {\bibfnamefont {W.}~\bibnamefont {Yan}},\
  }\href {\doibase 10.1038/s41467-021-22122-2} {\bibfield  {journal} {\bibinfo
  {journal} {Nature Communications}\ }\textbf {\bibinfo {volume} {12}},\
  \bibinfo {pages} {1854} (\bibinfo {year} {2021})}\BibitemShut {NoStop}%
\bibitem [{\citenamefont {Bykov}\ \emph {et~al.}(2021)\citenamefont {Bykov},
  \citenamefont {Fedotenko}, \citenamefont {Chariton}, \citenamefont {Laniel},
  \citenamefont {Glazyrin}, \citenamefont {Hanfland}, \citenamefont {Smith},
  \citenamefont {Prakapenka}, \citenamefont {Mahmood}, \citenamefont
  {Goncharov}, \citenamefont {Ponomareva}, \citenamefont {Tasn{\'{a}}di},
  \citenamefont {Abrikosov}, \citenamefont {{Bin Masood}}, \citenamefont
  {Hotz}, \citenamefont {Rudenko}, \citenamefont {Katsnelson}, \citenamefont
  {Dubrovinskaia}, \citenamefont {Dubrovinsky},\ and\ \citenamefont
  {Abrikosov}}]{Bykov2021}%
  \BibitemOpen
  \bibfield  {author} {\bibinfo {author} {\bibfnamefont {M.}~\bibnamefont
  {Bykov}}, \bibinfo {author} {\bibfnamefont {T.}~\bibnamefont {Fedotenko}},
  \bibinfo {author} {\bibfnamefont {S.}~\bibnamefont {Chariton}}, \bibinfo
  {author} {\bibfnamefont {D.}~\bibnamefont {Laniel}}, \bibinfo {author}
  {\bibfnamefont {K.}~\bibnamefont {Glazyrin}}, \bibinfo {author}
  {\bibfnamefont {M.}~\bibnamefont {Hanfland}}, \bibinfo {author}
  {\bibfnamefont {J.~S.}\ \bibnamefont {Smith}}, \bibinfo {author}
  {\bibfnamefont {V.~B.}\ \bibnamefont {Prakapenka}}, \bibinfo {author}
  {\bibfnamefont {M.~F.}\ \bibnamefont {Mahmood}}, \bibinfo {author}
  {\bibfnamefont {A.~F.}\ \bibnamefont {Goncharov}}, \bibinfo {author}
  {\bibfnamefont {A.~V.}\ \bibnamefont {Ponomareva}}, \bibinfo {author}
  {\bibfnamefont {F.}~\bibnamefont {Tasn{\'{a}}di}}, \bibinfo {author}
  {\bibfnamefont {A.~I.}\ \bibnamefont {Abrikosov}}, \bibinfo {author}
  {\bibfnamefont {T.}~\bibnamefont {{Bin Masood}}}, \bibinfo {author}
  {\bibfnamefont {I.}~\bibnamefont {Hotz}}, \bibinfo {author} {\bibfnamefont
  {A.~N.}\ \bibnamefont {Rudenko}}, \bibinfo {author} {\bibfnamefont {M.~I.}\
  \bibnamefont {Katsnelson}}, \bibinfo {author} {\bibfnamefont
  {N.}~\bibnamefont {Dubrovinskaia}}, \bibinfo {author} {\bibfnamefont
  {L.}~\bibnamefont {Dubrovinsky}}, \ and\ \bibinfo {author} {\bibfnamefont
  {I.~A.}\ \bibnamefont {Abrikosov}},\ }\href {\doibase
  10.1103/PhysRevLett.126.175501} {\bibfield  {journal} {\bibinfo  {journal}
  {Physical Review Letters}\ }\textbf {\bibinfo {volume} {126}},\ \bibinfo
  {pages} {175501} (\bibinfo {year} {2021})}\BibitemShut {NoStop}%
\bibitem [{\citenamefont {Liu}\ \emph {et~al.}(2021{\natexlab{a}})\citenamefont
  {Liu}, \citenamefont {Zhang}, \citenamefont {Gao},\ and\ \citenamefont
  {Yan}}]{Liu2021}%
  \BibitemOpen
  \bibfield  {author} {\bibinfo {author} {\bibfnamefont {D.}~\bibnamefont
  {Liu}}, \bibinfo {author} {\bibfnamefont {S.}~\bibnamefont {Zhang}}, \bibinfo
  {author} {\bibfnamefont {M.}~\bibnamefont {Gao}}, \ and\ \bibinfo {author}
  {\bibfnamefont {X.-W.}\ \bibnamefont {Yan}},\ }\href {\doibase
  10.1103/PhysRevB.103.125407} {\bibfield  {journal} {\bibinfo  {journal}
  {Physical Review B}\ }\textbf {\bibinfo {volume} {103}},\ \bibinfo {pages}
  {125407} (\bibinfo {year} {2021}{\natexlab{a}})}\BibitemShut {NoStop}%
\bibitem [{\citenamefont {Feng}\ \emph {et~al.}(2022)\citenamefont {Feng},
  \citenamefont {Zhang}, \citenamefont {Liu}, \citenamefont {Gao},
  \citenamefont {Ma}, \citenamefont {Yan},\ and\ \citenamefont
  {Xie}}]{Feng2022}%
  \BibitemOpen
  \bibfield  {author} {\bibinfo {author} {\bibfnamefont {P.}~\bibnamefont
  {Feng}}, \bibinfo {author} {\bibfnamefont {S.}~\bibnamefont {Zhang}},
  \bibinfo {author} {\bibfnamefont {D.}~\bibnamefont {Liu}}, \bibinfo {author}
  {\bibfnamefont {M.}~\bibnamefont {Gao}}, \bibinfo {author} {\bibfnamefont
  {F.}~\bibnamefont {Ma}}, \bibinfo {author} {\bibfnamefont {X.-W.}\
  \bibnamefont {Yan}}, \ and\ \bibinfo {author} {\bibfnamefont {Z.~Y.}\
  \bibnamefont {Xie}},\ }\href {\doibase 10.1021/acs.jpcc.2c02593} {\bibfield
  {journal} {\bibinfo  {journal} {The Journal of Physical Chemistry C}\
  }\textbf {\bibinfo {volume} {126}},\ \bibinfo {pages} {10139} (\bibinfo
  {year} {2022})}\BibitemShut {NoStop}%
\bibitem [{\citenamefont {Liu}\ \emph {et~al.}(2021{\natexlab{b}})\citenamefont
  {Liu}, \citenamefont {Feng}, \citenamefont {Gao},\ and\ \citenamefont
  {Yan}}]{Liu2021a}%
  \BibitemOpen
  \bibfield  {author} {\bibinfo {author} {\bibfnamefont {D.}~\bibnamefont
  {Liu}}, \bibinfo {author} {\bibfnamefont {P.}~\bibnamefont {Feng}}, \bibinfo
  {author} {\bibfnamefont {M.}~\bibnamefont {Gao}}, \ and\ \bibinfo {author}
  {\bibfnamefont {X.-W.}\ \bibnamefont {Yan}},\ }\href {\doibase
  10.1103/PhysRevB.103.155411} {\bibfield  {journal} {\bibinfo  {journal}
  {Physical Review B}\ }\textbf {\bibinfo {volume} {103}},\ \bibinfo {pages}
  {155411} (\bibinfo {year} {2021}{\natexlab{b}})}\BibitemShut {NoStop}%
\bibitem [{\citenamefont {Liu}\ \emph {et~al.}(2021{\natexlab{c}})\citenamefont
  {Liu}, \citenamefont {Zhang}, \citenamefont {Gao}, \citenamefont {Yan},\ and\
  \citenamefont {Xie}}]{Liu2021b}%
  \BibitemOpen
  \bibfield  {author} {\bibinfo {author} {\bibfnamefont {D.}~\bibnamefont
  {Liu}}, \bibinfo {author} {\bibfnamefont {S.}~\bibnamefont {Zhang}}, \bibinfo
  {author} {\bibfnamefont {M.}~\bibnamefont {Gao}}, \bibinfo {author}
  {\bibfnamefont {X.-W.}\ \bibnamefont {Yan}}, \ and\ \bibinfo {author}
  {\bibfnamefont {Z.~Y.}\ \bibnamefont {Xie}},\ }\href {\doibase
  10.1063/5.0054730} {\bibfield  {journal} {\bibinfo  {journal} {Applied
  Physics Letters}\ }\textbf {\bibinfo {volume} {118}},\ \bibinfo {pages}
  {223104} (\bibinfo {year} {2021}{\natexlab{c}})}\BibitemShut {NoStop}%
\bibitem [{\citenamefont {Wang}\ \emph {et~al.}(2021)\citenamefont {Wang},
  \citenamefont {Niu},\ and\ \citenamefont {Qiao}}]{Wang2021}%
  \BibitemOpen
  \bibfield  {author} {\bibinfo {author} {\bibfnamefont {H.}~\bibnamefont
  {Wang}}, \bibinfo {author} {\bibfnamefont {Q.}~\bibnamefont {Niu}}, \ and\
  \bibinfo {author} {\bibfnamefont {Z.}~\bibnamefont {Qiao}},\ }\href {\doibase
  10.1103/PhysRevB.104.235157} {\bibfield  {journal} {\bibinfo  {journal}
  {Physical Review B}\ }\textbf {\bibinfo {volume} {104}},\ \bibinfo {pages}
  {235157} (\bibinfo {year} {2021})}\BibitemShut {NoStop}%
\bibitem [{\citenamefont {Dong}\ \emph {et~al.}(2022)\citenamefont {Dong},
  \citenamefont {Wang}, \citenamefont {Zhao}, \citenamefont {Gao},
  \citenamefont {Yan}, \citenamefont {Ma},\ and\ \citenamefont
  {Lu}}]{Dong2022}%
  \BibitemOpen
  \bibfield  {author} {\bibinfo {author} {\bibfnamefont {J.}~\bibnamefont
  {Dong}}, \bibinfo {author} {\bibfnamefont {C.}~\bibnamefont {Wang}}, \bibinfo
  {author} {\bibfnamefont {X.}~\bibnamefont {Zhao}}, \bibinfo {author}
  {\bibfnamefont {M.}~\bibnamefont {Gao}}, \bibinfo {author} {\bibfnamefont
  {X.-W.}\ \bibnamefont {Yan}}, \bibinfo {author} {\bibfnamefont
  {F.}~\bibnamefont {Ma}}, \ and\ \bibinfo {author} {\bibfnamefont {Z.-Y.}\
  \bibnamefont {Lu}},\ }\href {\doibase 10.1103/PhysRevMaterials.6.074202}
  {\bibfield  {journal} {\bibinfo  {journal} {Physical Review Materials}\
  }\textbf {\bibinfo {volume} {6}},\ \bibinfo {pages} {074202} (\bibinfo {year}
  {2022})}\BibitemShut {NoStop}%
\bibitem [{\citenamefont {Zhang}\ \emph {et~al.}(2015)\citenamefont {Zhang},
  \citenamefont {Li}, \citenamefont {Zhao},\ and\ \citenamefont
  {Wang}}]{Zhang2015}%
  \BibitemOpen
  \bibfield  {author} {\bibinfo {author} {\bibfnamefont {S.}~\bibnamefont
  {Zhang}}, \bibinfo {author} {\bibfnamefont {Y.}~\bibnamefont {Li}}, \bibinfo
  {author} {\bibfnamefont {T.}~\bibnamefont {Zhao}}, \ and\ \bibinfo {author}
  {\bibfnamefont {Q.}~\bibnamefont {Wang}},\ }\href {\doibase
  10.1038/srep05241} {\bibfield  {journal} {\bibinfo  {journal} {Scientific
  Reports}\ }\textbf {\bibinfo {volume} {4}},\ \bibinfo {pages} {5241}
  (\bibinfo {year} {2015})}\BibitemShut {NoStop}%
\bibitem [{\citenamefont {Zhao}\ and\ \citenamefont {Wang}(2020)}]{Zhao2020}%
  \BibitemOpen
  \bibfield  {author} {\bibinfo {author} {\bibfnamefont {K.}~\bibnamefont
  {Zhao}}\ and\ \bibinfo {author} {\bibfnamefont {Q.}~\bibnamefont {Wang}},\
  }\href {\doibase 10.1016/j.apsusc.2019.144620} {\bibfield  {journal}
  {\bibinfo  {journal} {Applied Surface Science}\ }\textbf {\bibinfo {volume}
  {505}},\ \bibinfo {pages} {144620} (\bibinfo {year} {2020})}\BibitemShut
  {NoStop}%
\bibitem [{\citenamefont {Mortazavi}\ \emph {et~al.}(2021)\citenamefont
  {Mortazavi}, \citenamefont {Shojaei},\ and\ \citenamefont
  {Zhuang}}]{Mortazavi2021}%
  \BibitemOpen
  \bibfield  {author} {\bibinfo {author} {\bibfnamefont {B.}~\bibnamefont
  {Mortazavi}}, \bibinfo {author} {\bibfnamefont {F.}~\bibnamefont {Shojaei}},
  \ and\ \bibinfo {author} {\bibfnamefont {X.}~\bibnamefont {Zhuang}},\ }\href
  {\doibase 10.1016/j.mtnano.2021.100125} {\bibfield  {journal} {\bibinfo
  {journal} {Materials Today Nano}\ }\textbf {\bibinfo {volume} {15}},\
  \bibinfo {pages} {2} (\bibinfo {year} {2021})}\BibitemShut {NoStop}%
\bibitem [{\citenamefont {Kresse}\ and\ \citenamefont
  {Hafner}(1993)}]{PhysRevB.47.558}%
  \BibitemOpen
  \bibfield  {author} {\bibinfo {author} {\bibfnamefont {G.}~\bibnamefont
  {Kresse}}\ and\ \bibinfo {author} {\bibfnamefont {J.}~\bibnamefont
  {Hafner}},\ }\href {\doibase 10.1103/PhysRevB.47.558} {\bibfield  {journal}
  {\bibinfo  {journal} {Physical Review B}\ }\textbf {\bibinfo {volume} {47}},\
  \bibinfo {pages} {558} (\bibinfo {year} {1993})}\BibitemShut {NoStop}%
\bibitem [{\citenamefont {Kresse}\ and\ \citenamefont
  {Furthm{\"{u}}ller}(1996)}]{PhysRevB.54.11169}%
  \BibitemOpen
  \bibfield  {author} {\bibinfo {author} {\bibfnamefont {G.}~\bibnamefont
  {Kresse}}\ and\ \bibinfo {author} {\bibfnamefont {J.}~\bibnamefont
  {Furthm{\"{u}}ller}},\ }\href {\doibase 10.1103/PhysRevB.54.11169} {\bibfield
   {journal} {\bibinfo  {journal} {Physical Review B}\ }\textbf {\bibinfo
  {volume} {54}},\ \bibinfo {pages} {11169} (\bibinfo {year}
  {1996})}\BibitemShut {NoStop}%
\bibitem [{\citenamefont {Perdew}\ \emph {et~al.}(1996)\citenamefont {Perdew},
  \citenamefont {Burke},\ and\ \citenamefont
  {Ernzerhof}}]{PhysRevLett.77.3865}%
  \BibitemOpen
  \bibfield  {author} {\bibinfo {author} {\bibfnamefont {J.~P.}\ \bibnamefont
  {Perdew}}, \bibinfo {author} {\bibfnamefont {K.}~\bibnamefont {Burke}}, \
  and\ \bibinfo {author} {\bibfnamefont {M.}~\bibnamefont {Ernzerhof}},\ }\href
  {\doibase 10.1103/PhysRevLett.77.3865} {\bibfield  {journal} {\bibinfo
  {journal} {Physical Review Letters}\ }\textbf {\bibinfo {volume} {77}},\
  \bibinfo {pages} {3865} (\bibinfo {year} {1996})}\BibitemShut {NoStop}%
\bibitem [{\citenamefont {Bl{\"{o}}chl}(1994)}]{PhysRevB.50.17953}%
  \BibitemOpen
  \bibfield  {author} {\bibinfo {author} {\bibfnamefont {P.~E.}\ \bibnamefont
  {Bl{\"{o}}chl}},\ }\href {\doibase 10.1103/PhysRevB.50.17953} {\bibfield
  {journal} {\bibinfo  {journal} {Physical Review B}\ }\textbf {\bibinfo
  {volume} {50}},\ \bibinfo {pages} {17953} (\bibinfo {year}
  {1994})}\BibitemShut {NoStop}%
\bibitem [{\citenamefont {Togo}\ and\ \citenamefont {Tanaka}(2015)}]{Togo2015}%
  \BibitemOpen
  \bibfield  {author} {\bibinfo {author} {\bibfnamefont {A.}~\bibnamefont
  {Togo}}\ and\ \bibinfo {author} {\bibfnamefont {I.}~\bibnamefont {Tanaka}},\
  }\href {\doibase 10.1016/j.scriptamat.2015.07.021} {\bibfield  {journal}
  {\bibinfo  {journal} {Scripta Materialia}\ }\textbf {\bibinfo {volume}
  {108}},\ \bibinfo {pages} {1} (\bibinfo {year} {2015})}\BibitemShut {NoStop}%
\bibitem [{\citenamefont {Martyna}\ \emph {et~al.}(1992)\citenamefont
  {Martyna}, \citenamefont {Klein},\ and\ \citenamefont
  {Tuckerman}}]{Martyna1992}%
  \BibitemOpen
  \bibfield  {author} {\bibinfo {author} {\bibfnamefont {G.~J.}\ \bibnamefont
  {Martyna}}, \bibinfo {author} {\bibfnamefont {M.~L.}\ \bibnamefont {Klein}},
  \ and\ \bibinfo {author} {\bibfnamefont {M.}~\bibnamefont {Tuckerman}},\
  }\href {\doibase 10.1063/1.463940} {\bibfield  {journal} {\bibinfo  {journal}
  {The Journal of Chemical Physics}\ }\textbf {\bibinfo {volume} {97}},\
  \bibinfo {pages} {2635} (\bibinfo {year} {1992})}\BibitemShut {NoStop}%
\bibitem [{\citenamefont {Zhang}\ \emph {et~al.}(2021)\citenamefont {Zhang},
  \citenamefont {Wang}, \citenamefont {Guo}, \citenamefont {Li},\ and\
  \citenamefont {Wang}}]{Zhang2021}%
  \BibitemOpen
  \bibfield  {author} {\bibinfo {author} {\bibfnamefont {Y.}~\bibnamefont
  {Zhang}}, \bibinfo {author} {\bibfnamefont {B.}~\bibnamefont {Wang}},
  \bibinfo {author} {\bibfnamefont {Y.}~\bibnamefont {Guo}}, \bibinfo {author}
  {\bibfnamefont {Q.}~\bibnamefont {Li}}, \ and\ \bibinfo {author}
  {\bibfnamefont {J.}~\bibnamefont {Wang}},\ }\href {\doibase
  10.1016/j.commatsci.2021.110638} {\bibfield  {journal} {\bibinfo  {journal}
  {Computational Materials Science}\ }\textbf {\bibinfo {volume} {197}},\
  \bibinfo {pages} {110638} (\bibinfo {year} {2021})}\BibitemShut {NoStop}%
\bibitem [{\citenamefont {Sun}\ \emph {et~al.}(2015)\citenamefont {Sun},
  \citenamefont {Ruzsinszky},\ and\ \citenamefont {Perdew}}]{Sun2015}%
  \BibitemOpen
  \bibfield  {author} {\bibinfo {author} {\bibfnamefont {J.}~\bibnamefont
  {Sun}}, \bibinfo {author} {\bibfnamefont {A.}~\bibnamefont {Ruzsinszky}}, \
  and\ \bibinfo {author} {\bibfnamefont {J.~P.}\ \bibnamefont {Perdew}},\
  }\href {\doibase 10.1103/PhysRevLett.115.036402} {\bibfield  {journal}
  {\bibinfo  {journal} {Physical Review Letters}\ }\textbf {\bibinfo {volume}
  {115}},\ \bibinfo {pages} {036402} (\bibinfo {year} {2015})}\BibitemShut
  {NoStop}%
\bibitem [{\citenamefont {Fan}\ \emph {et~al.}(2021)\citenamefont {Fan},
  \citenamefont {Yan}, \citenamefont {Tripp}, \citenamefont
  {Krej{\v{c}}{\'{i}}}, \citenamefont {Dimosthenous}, \citenamefont {Kachel},
  \citenamefont {Chen}, \citenamefont {Foster}, \citenamefont {Koert},
  \citenamefont {Liljeroth},\ and\ \citenamefont {Gottfried}}]{Fan2021}%
  \BibitemOpen
  \bibfield  {author} {\bibinfo {author} {\bibfnamefont {Q.}~\bibnamefont
  {Fan}}, \bibinfo {author} {\bibfnamefont {L.}~\bibnamefont {Yan}}, \bibinfo
  {author} {\bibfnamefont {M.~W.}\ \bibnamefont {Tripp}}, \bibinfo {author}
  {\bibfnamefont {O.}~\bibnamefont {Krej{\v{c}}{\'{i}}}}, \bibinfo {author}
  {\bibfnamefont {S.}~\bibnamefont {Dimosthenous}}, \bibinfo {author}
  {\bibfnamefont {S.~R.}\ \bibnamefont {Kachel}}, \bibinfo {author}
  {\bibfnamefont {M.}~\bibnamefont {Chen}}, \bibinfo {author} {\bibfnamefont
  {A.~S.}\ \bibnamefont {Foster}}, \bibinfo {author} {\bibfnamefont
  {U.}~\bibnamefont {Koert}}, \bibinfo {author} {\bibfnamefont
  {P.}~\bibnamefont {Liljeroth}}, \ and\ \bibinfo {author} {\bibfnamefont
  {J.~M.}\ \bibnamefont {Gottfried}},\ }\href {\doibase
  10.1126/science.abg4509} {\bibfield  {journal} {\bibinfo  {journal}
  {Science}\ }\textbf {\bibinfo {volume} {372}},\ \bibinfo {pages} {852}
  (\bibinfo {year} {2021})}\BibitemShut {NoStop}%
\bibitem [{\citenamefont {Wilsdorf}(1948)}]{Wilsdorf1948}%
  \BibitemOpen
  \bibfield  {author} {\bibinfo {author} {\bibfnamefont {H.}~\bibnamefont
  {Wilsdorf}},\ }\href {\doibase 10.1107/S0365110X48000314} {\bibfield
  {journal} {\bibinfo  {journal} {Acta Crystallographica}\ }\textbf {\bibinfo
  {volume} {1}},\ \bibinfo {pages} {115} (\bibinfo {year} {1948})}\BibitemShut
  {NoStop}%
\bibitem [{\citenamefont {Crowhurst}(2006)}]{Crowhurst2006}%
  \BibitemOpen
  \bibfield  {author} {\bibinfo {author} {\bibfnamefont {J.~C.}\ \bibnamefont
  {Crowhurst}},\ }\href {\doibase 10.1126/science.1121813} {\bibfield
  {journal} {\bibinfo  {journal} {Science}\ }\textbf {\bibinfo {volume}
  {311}},\ \bibinfo {pages} {1275} (\bibinfo {year} {2006})}\BibitemShut
  {NoStop}%
\bibitem [{\citenamefont {Groenewolt}\ and\ \citenamefont
  {Antonietti}(2005)}]{Groenewolt2005}%
  \BibitemOpen
  \bibfield  {author} {\bibinfo {author} {\bibfnamefont {M.}~\bibnamefont
  {Groenewolt}}\ and\ \bibinfo {author} {\bibfnamefont {M.}~\bibnamefont
  {Antonietti}},\ }\href {\doibase 10.1002/adma.200401756} {\bibfield
  {journal} {\bibinfo  {journal} {Advanced Materials}\ }\textbf {\bibinfo
  {volume} {17}},\ \bibinfo {pages} {1789} (\bibinfo {year}
  {2005})}\BibitemShut {NoStop}%
\bibitem [{\citenamefont {Kroll}\ \emph {et~al.}(2012)\citenamefont {Kroll},
  \citenamefont {Kraus}, \citenamefont {Sch{\"{o}}nfelder}, \citenamefont
  {Aristov}, \citenamefont {Molodtsova}, \citenamefont {Hoffmann},\ and\
  \citenamefont {Knupfer}}]{Kroll2012}%
  \BibitemOpen
  \bibfield  {author} {\bibinfo {author} {\bibfnamefont {T.}~\bibnamefont
  {Kroll}}, \bibinfo {author} {\bibfnamefont {R.}~\bibnamefont {Kraus}},
  \bibinfo {author} {\bibfnamefont {R.}~\bibnamefont {Sch{\"{o}}nfelder}},
  \bibinfo {author} {\bibfnamefont {V.~Y.}\ \bibnamefont {Aristov}}, \bibinfo
  {author} {\bibfnamefont {O.~V.}\ \bibnamefont {Molodtsova}}, \bibinfo
  {author} {\bibfnamefont {P.}~\bibnamefont {Hoffmann}}, \ and\ \bibinfo
  {author} {\bibfnamefont {M.}~\bibnamefont {Knupfer}},\ }\href {\doibase
  10.1063/1.4738754} {\bibfield  {journal} {\bibinfo  {journal} {The Journal of
  Chemical Physics}\ }\textbf {\bibinfo {volume} {137}},\ \bibinfo {pages}
  {054306} (\bibinfo {year} {2012})}\BibitemShut {NoStop}%
\bibitem [{\citenamefont {Chen}\ \emph {et~al.}(2015)\citenamefont {Chen},
  \citenamefont {Dai},\ and\ \citenamefont {Zeng}}]{Chen2015}%
  \BibitemOpen
  \bibfield  {author} {\bibinfo {author} {\bibfnamefont {S.}~\bibnamefont
  {Chen}}, \bibinfo {author} {\bibfnamefont {J.}~\bibnamefont {Dai}}, \ and\
  \bibinfo {author} {\bibfnamefont {X.~C.}\ \bibnamefont {Zeng}},\ }\href
  {\doibase 10.1039/C4CP05328A} {\bibfield  {journal} {\bibinfo  {journal}
  {Physical Chemistry Chemical Physics}\ }\textbf {\bibinfo {volume} {17}},\
  \bibinfo {pages} {5954} (\bibinfo {year} {2015})}\BibitemShut {NoStop}%
\bibitem [{\citenamefont {Huang}\ \emph {et~al.}(2017)\citenamefont {Huang},
  \citenamefont {Clark}, \citenamefont {Navarro-Moratalla}, \citenamefont
  {Klein}, \citenamefont {Cheng}, \citenamefont {Seyler}, \citenamefont
  {Zhong}, \citenamefont {Schmidgall}, \citenamefont {McGuire}, \citenamefont
  {Cobden}, \citenamefont {Yao}, \citenamefont {Xiao}, \citenamefont
  {Jarillo-Herrero},\ and\ \citenamefont {Xu}}]{Huang2017}%
  \BibitemOpen
  \bibfield  {author} {\bibinfo {author} {\bibfnamefont {B.}~\bibnamefont
  {Huang}}, \bibinfo {author} {\bibfnamefont {G.}~\bibnamefont {Clark}},
  \bibinfo {author} {\bibfnamefont {E.}~\bibnamefont {Navarro-Moratalla}},
  \bibinfo {author} {\bibfnamefont {D.~R.}\ \bibnamefont {Klein}}, \bibinfo
  {author} {\bibfnamefont {R.}~\bibnamefont {Cheng}}, \bibinfo {author}
  {\bibfnamefont {K.~L.}\ \bibnamefont {Seyler}}, \bibinfo {author}
  {\bibfnamefont {D.}~\bibnamefont {Zhong}}, \bibinfo {author} {\bibfnamefont
  {E.}~\bibnamefont {Schmidgall}}, \bibinfo {author} {\bibfnamefont {M.~A.}\
  \bibnamefont {McGuire}}, \bibinfo {author} {\bibfnamefont {D.~H.}\
  \bibnamefont {Cobden}}, \bibinfo {author} {\bibfnamefont {W.}~\bibnamefont
  {Yao}}, \bibinfo {author} {\bibfnamefont {D.}~\bibnamefont {Xiao}}, \bibinfo
  {author} {\bibfnamefont {P.}~\bibnamefont {Jarillo-Herrero}}, \ and\ \bibinfo
  {author} {\bibfnamefont {X.}~\bibnamefont {Xu}},\ }\href {\doibase
  10.1038/nature22391} {\bibfield  {journal} {\bibinfo  {journal} {Nature}\
  }\textbf {\bibinfo {volume} {546}},\ \bibinfo {pages} {270} (\bibinfo {year}
  {2017})}\BibitemShut {NoStop}%
\bibitem [{\citenamefont {Burch}\ \emph {et~al.}(2018)\citenamefont {Burch},
  \citenamefont {Mandrus},\ and\ \citenamefont {Park}}]{Burch2018}%
  \BibitemOpen
  \bibfield  {author} {\bibinfo {author} {\bibfnamefont {K.~S.}\ \bibnamefont
  {Burch}}, \bibinfo {author} {\bibfnamefont {D.}~\bibnamefont {Mandrus}}, \
  and\ \bibinfo {author} {\bibfnamefont {J.~G.}\ \bibnamefont {Park}},\ }\href
  {\doibase 10.1038/s41586-018-0631-z} {\bibfield  {journal} {\bibinfo
  {journal} {Nature}\ }\textbf {\bibinfo {volume} {563}},\ \bibinfo {pages}
  {47} (\bibinfo {year} {2018})}\BibitemShut {NoStop}%
\bibitem [{\citenamefont {Cortie}\ \emph {et~al.}(2020)\citenamefont {Cortie},
  \citenamefont {Causer}, \citenamefont {Rule}, \citenamefont {Fritzsche},
  \citenamefont {Kreuzpaintner},\ and\ \citenamefont {Klose}}]{Cortie2020}%
  \BibitemOpen
  \bibfield  {author} {\bibinfo {author} {\bibfnamefont {D.~L.}\ \bibnamefont
  {Cortie}}, \bibinfo {author} {\bibfnamefont {G.~L.}\ \bibnamefont {Causer}},
  \bibinfo {author} {\bibfnamefont {K.~C.}\ \bibnamefont {Rule}}, \bibinfo
  {author} {\bibfnamefont {H.}~\bibnamefont {Fritzsche}}, \bibinfo {author}
  {\bibfnamefont {W.}~\bibnamefont {Kreuzpaintner}}, \ and\ \bibinfo {author}
  {\bibfnamefont {F.}~\bibnamefont {Klose}},\ }\href {\doibase
  10.1002/adfm.201901414} {\bibfield  {journal} {\bibinfo  {journal} {Advanced
  Functional Materials}\ }\textbf {\bibinfo {volume} {30}},\ \bibinfo {pages}
  {1901414} (\bibinfo {year} {2020})}\BibitemShut {NoStop}%
\bibitem [{\citenamefont {Ma}\ \emph {et~al.}(2009)\citenamefont {Ma},
  \citenamefont {Ji}, \citenamefont {Hu}, \citenamefont {Lu},\ and\
  \citenamefont {Xiang}}]{Ma2009}%
  \BibitemOpen
  \bibfield  {author} {\bibinfo {author} {\bibfnamefont {F.}~\bibnamefont
  {Ma}}, \bibinfo {author} {\bibfnamefont {W.}~\bibnamefont {Ji}}, \bibinfo
  {author} {\bibfnamefont {J.}~\bibnamefont {Hu}}, \bibinfo {author}
  {\bibfnamefont {Z.-Y.}\ \bibnamefont {Lu}}, \ and\ \bibinfo {author}
  {\bibfnamefont {T.}~\bibnamefont {Xiang}},\ }\href {\doibase
  10.1103/PhysRevLett.102.177003} {\bibfield  {journal} {\bibinfo  {journal}
  {Physical Review Letters}\ }\textbf {\bibinfo {volume} {102}},\ \bibinfo
  {pages} {177003} (\bibinfo {year} {2009})}\BibitemShut {NoStop}%
\bibitem [{\citenamefont {Yu}\ and\ \citenamefont {Si}(2015)}]{Yu2015}%
  \BibitemOpen
  \bibfield  {author} {\bibinfo {author} {\bibfnamefont {R.}~\bibnamefont
  {Yu}}\ and\ \bibinfo {author} {\bibfnamefont {Q.}~\bibnamefont {Si}},\ }\href
  {\doibase 10.1103/PhysRevLett.115.116401} {\bibfield  {journal} {\bibinfo
  {journal} {Physical Review Letters}\ }\textbf {\bibinfo {volume} {115}},\
  \bibinfo {pages} {116401} (\bibinfo {year} {2015})}\BibitemShut {NoStop}%
\bibitem [{\citenamefont {Zhao}\ \emph {et~al.}(2009)\citenamefont {Zhao},
  \citenamefont {Adroja}, \citenamefont {Yao}, \citenamefont {Bewley},
  \citenamefont {Li}, \citenamefont {Wang}, \citenamefont {Wu}, \citenamefont
  {Chen}, \citenamefont {Hu},\ and\ \citenamefont {Dai}}]{Zhao2009}%
  \BibitemOpen
  \bibfield  {author} {\bibinfo {author} {\bibfnamefont {J.}~\bibnamefont
  {Zhao}}, \bibinfo {author} {\bibfnamefont {D.~T.}\ \bibnamefont {Adroja}},
  \bibinfo {author} {\bibfnamefont {D.~X.}\ \bibnamefont {Yao}}, \bibinfo
  {author} {\bibfnamefont {R.}~\bibnamefont {Bewley}}, \bibinfo {author}
  {\bibfnamefont {S.}~\bibnamefont {Li}}, \bibinfo {author} {\bibfnamefont
  {X.~F.}\ \bibnamefont {Wang}}, \bibinfo {author} {\bibfnamefont
  {G.}~\bibnamefont {Wu}}, \bibinfo {author} {\bibfnamefont {X.~H.}\
  \bibnamefont {Chen}}, \bibinfo {author} {\bibfnamefont {J.}~\bibnamefont
  {Hu}}, \ and\ \bibinfo {author} {\bibfnamefont {P.}~\bibnamefont {Dai}},\
  }\href {\doibase 10.1038/nphys1336} {\bibfield  {journal} {\bibinfo
  {journal} {Nature Physics}\ }\textbf {\bibinfo {volume} {5}},\ \bibinfo
  {pages} {555} (\bibinfo {year} {2009})}\BibitemShut {NoStop}%
\end{thebibliography}%

\end{document}